\documentclass[12pt, onecolumn, letterpaper]{article} 

\usepackage[utf8]{inputenc} 
\usepackage[T1]{fontenc}
\usepackage{url}
\usepackage{ifthen}
\usepackage{cite}
\usepackage[cmex10]{amsmath} 
\usepackage{amssymb}
\usepackage{algorithm}
\usepackage{algorithmic}
\usepackage{subfigure}
\usepackage{tikz}
\usepackage{algorithm}
\usepackage{algorithmic}
\usepackage{amsmath}
\usepackage{amssymb}
\usepackage{ecltree}
\usepackage{enumerate}
\usepackage{mathrsfs}
\usepackage{xcolor}
\usepackage{comment}
\usepackage{tikz}
\usepackage{wrapfig}
\usetikzlibrary{trees}

\usepackage{here}
\usepackage{cases}
\usepackage{stackengine}
\usepackage{color}
\usepackage[normalem]{ulem}
\usepackage{longtable}

\def\BibTeX{{\rm B\kern-.05em{\sc i\kern-.025em b}\kern-.08em
    T\kern-.1667em\lower.7ex\hbox{E}\kern-.125emX}}

\newtheorem{definition}{Definition}
\newtheorem{lemma}{Lemma}

\newtheorem{theorem}{Theorem}

\newtheorem{example}{Example}

\def\qed{\hfill $\Box$}

\def\trans{\mathrm{\tau}}

\def\PREF{\mathcal{P}}

\def\hdec{\text{-}\mathrm{dec}}

\def\reg{\mathrm{reg}}

\def\fork{\mathrm{fork}}
\def\ext{\mathrm{ext}}
\def\Huff{\mathrm{Huff}}

\def\suff{\mathrm{suf}}

\newcommand{\argmin}{\mathop{\rm arg~min}\limits}
\newcommand{\succpreceq}{\mathrel{\vcenter{\hbox{$\stackMath\stackinset{l}{-0.28em}{c}{1.1ex}{\prec}{\succeq}$}}}}

\usepackage{setspace}
\begin{document}

\title{Optimality of Huffman Code in the Class of $1$-bit Delay Decodable Codes}
\author{Kengo Hashimoto, Ken-ichi Iwata}     

\date{University of Fukui, \quad
             E-mail: \{khasimot, k-iwata\}@u-fukui.ac.jp}
\maketitle

\begin{abstract}
For a given independent and identically distributed (i.i.d.) source, Huffman code achieves the optimal average codeword length in the class of instantaneous code with a single code table.
However, it is known that there exist time-variant encoders, which achieve a shorter average codeword length than the Huffman code, using multiple code tables and allowing at most $k$-bit decoding delay for $k = 2, 3, 4, \ldots $.
On the other hand, it is not known whether there exists a $1$-bit delay decodable code, which achieves a shorter average length than the Huffman code.
This paper proves that for a given i.i.d.\ source, a Huffman code achieves the optimal average codeword length in the class of $1$-bit delay decodable codes with a finite number of code tables.
\end{abstract}


\section{Introduction}
\label{sec:introduction}

\begin{figure}[H]
  \centering
  \includegraphics[keepaspectratio,width=130mm]{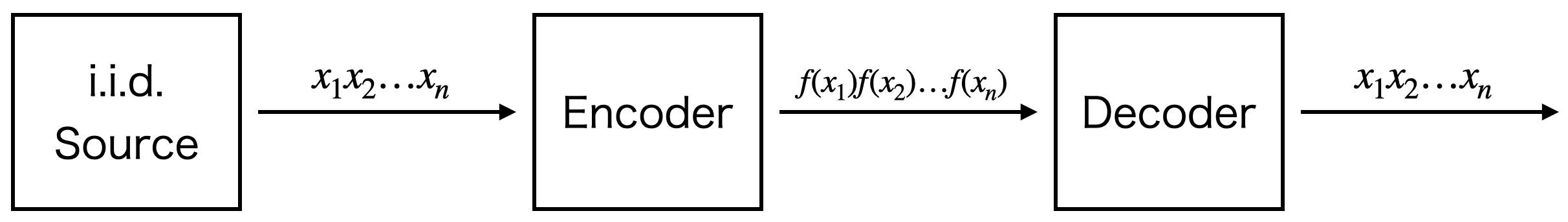}
      \vspace{-8pt}
  \caption{The data compression system in this paper.}
  \label{fig:system}
\end{figure}

We consider the data compression system shown in Fig.\ \ref{fig:system}.
The i.i.d.\ source outputs a sequence $\pmb{x} = x_1x_2\ldots x_n$ of symbols of the source alphabet $\mathcal{S} = \{s_1, s_2, \ldots, s_{\sigma}\}$, where $n$ and $\sigma$ denote the length of $\pmb{x}$ and the alphabet size, respectively.
Each source output follows a fixed probability distribution $(\mu(s_1), \mu(s_2), \ldots, \mu(s_{\sigma}))$, where $\mu(s_i)$ is the probability of occurrence of $s_i$ for $i = 1, 2, \ldots, \sigma$.
In this paper, we assume $\sigma \geq 2$.
The encoder reads the source sequence $\pmb{x}$ symbol by symbol from the beginning of $\pmb{x}$ and encodes them according to a code table $f : \mathcal{S} \rightarrow \mathcal{C}^{\ast}$, where $\mathcal{C}^{\ast}$ denotes the set of all sequences of finite length over $\mathcal{C}$.
As the result, $\pmb{x} = x_1x_2\ldots x_n$ is encoded to a binary codeword sequence $f(\pmb{x}) = f(x_1)f(x_2)\ldots f(x_n)$ over the coding alphabet $\mathcal{C} := \{0, 1\}$. 
Then the decoder reads the codeword sequence $f(\pmb{x})$ bit by bit from the beginning of $f(\pmb{x})$ and recovers the original source sequence $\pmb{x}$.

For example, consider an i.i.d. source with source alphabet $\mathcal{S} = \{\mathrm{a}, \mathrm{b}, \mathrm{c}, \mathrm{d}\}$,
a probability distribution $(\mu(\mathrm{a}), \mu(\mathrm{b}), \mu(\mathrm{c}), \mu(\mathrm{d})) = (0.1, 0.2, 0.3, 0.4)$ and the code table $f^{(\mathrm{I})}$ in Table \ref{tab:code-Huff}. The \emph{average codeword length} (per symbol) $L(f^{(\mathrm{I})})$ of the code table $f^{(\mathrm{I})}$ is calculated as
\begin{eqnarray}		
L(f^{(\mathrm{I})}) &=& \mu(\mathrm{a}) |f^{(\mathrm{I})}(\mathrm{a})| + \mu(\mathrm{b}) |f^{(\mathrm{I})}(\mathrm{b})| + \mu(\mathrm{c}) |f^{(\mathrm{I})}(\mathrm{c})| + \mu(\mathrm{d}) |f^{(\mathrm{I})}(\mathrm{d})| = 1.9 \label{eq:rrg7p2o0qoyd},
\end{eqnarray}
where $|\cdot|$ denotes the length of a sequence.

A source sequence $\pmb{x} = x_1x_2x_3x_4x_5 = \mathrm{adbac}$ is encoded to $f^{(\mathrm{I})}(\pmb{x}) = f^{(\mathrm{I})}(\mathrm{a})f^{(\mathrm{I})}(\mathrm{d})f^{(\mathrm{I})}(\mathrm{b})f^{(\mathrm{I})}(\mathrm{a})\allowbreak f^{(\mathrm{I})}(\mathrm{c}) = 000100100001$. 
The code table $f^{(\mathrm{I})}$ is \emph{prefix-free}; that is, for any two distinct symbols $s$ and $s'$, $f(s)$ is not a prefix of $f(s')$.
Therefore, in the decoding process, the decoder can uniquely determine $x_1 = \mathrm{a}$ immediately after reading the prefix $f^{(\mathrm{I})}(\mathrm{a}) = 000$ of $f^{(\mathrm{I})}(\pmb{x})$.
Also, the decoder can uniquely determine $x_1x_2 = \mathrm{a}\mathrm{d}$ immediately after reading the prefix $f^{(\mathrm{I})}(\mathrm{a})f^{(\mathrm{I})}(\mathrm{d}) = 0001$ of $f^{(\mathrm{I})}(\pmb{x})$.
In general, a code table $f$ is called an \emph{instantaneous code} if for any source sequence $\pmb{x} = x_1x_2\ldots x_n$ and its any prefix $x_1x_2\ldots x_l$, the decoder can uniquely recover $x_1x_2\ldots x_l$ immediately after reading the prefix $f(x_1)f(x_2)\ldots f(x_l)$.
In fact, the code table $f^{(\mathrm{I})}$ is an instantaneous code.

\begin{table}
\caption{An example of a code table.}
\label{tab:code-Huff}
\centering
\begin{tabular}{cl}
\hline
$s \in \mathcal{S}$ & $f^{(\mathrm{I})}(s)$\\
\hline \hline
a & 000\\
b & 001\\
c & 01\\
d & 1\\
\hline
\end{tabular}
\end{table}

Huffman proposed an algorithm to construct an instantaneous code that achieved the optimal (minimum) average codeword length for a given probability distribution $(\mu(s_1), \mu(s_2), \ldots, \allowbreak \mu(s_{\sigma}))$ \cite{Huffman1952}.
A code table constructed by Huffman's algorithm is called a \emph{Huffman code}.
Namely, a Huffman code achieves the optimal average codeword length in the class of instantaneous codes.
McMillan's theorem\cite{McMillan1956} implies that Huffman code is optimal also in the class of uniquely decodable codes (with a single code table).
The code table $f^{(\mathrm{I})}$ in Table \ref{tab:code-Huff} is a Huffman code for the probability distribution $(\mu(\mathrm{a}), \mu(\mathrm{b}), \mu(\mathrm{c}), \mu(\mathrm{d})) = (0.1, 0.2, 0.3, 0.4)$.

In 2015, Yamamoto, Tsuchihashi, and Honda\cite{Yamamoto2015} proposed binary AIFV code that can achieve a shorter average codeword length than the Huffman code.
An AIFV code uses a time-variant encoder consisting of two code tables $f_0, f_1$ and allows at most $2$-bit delay for decoding.
We omit details of the definition of AIFV code here and show an example of an AIFV code $F^{(\mathrm{II})}$ in Table \ref{tab:code-aifv}.
In the encoding process of $\pmb{x} = x_1x_2\ldots x_n$ with the AIFV code $F^{(\mathrm{II})}$, the first symbol $x_1$ is encoded with the code table $f_0$.
For $i = 2, 3, \ldots, n$, if $x_{i-1}$ is encoded with the code table $f_j$, then $x_i$ is encoded with the code table $f_{\trans_j(x_{i-1})}$.
For example, a source sequence $\pmb{x} = x_1x_2x_3x_4x_5 = \mathrm{adbac}$ is encoded to $f_0(\mathrm{a})f_0(\mathrm{d})f_1(\mathrm{b})f_1(\mathrm{a})f_0(\mathrm{c}) = 100111110001$. 

AIFV code $F^{(\mathrm{II})}$ is not an instantaneous code because the decoder cannot uniquely determine whether $x_1x_2 = \mathrm{a}\mathrm{d}$ or not at the time of reading $f_0(\mathrm{a})f_0(\mathrm{d}) = 1001$ (there are still two possibilities, $\pmb{x} = \mathrm{aa}\ldots$ and $\pmb{x} = \mathrm{ad}\ldots$).
However, the decoder can distinguish them by reading the following two bits; that is, the decoder can uniquely determine $x_1x_2 = \mathrm{a}\mathrm{d}$ immediately after reading $f_0(\mathrm{a})f_0(\mathrm{d})11 = 100111$.
Similarly, when the decoder reads the prefix $f_0(\mathrm{a})f_0(\mathrm{d})f_1(\mathrm{b})f_1(\mathrm{b}) = 10011111$, the decoder cannot uniquely determine whether $x_1x_2x_3x_4 = \mathrm{a}\mathrm{d}\mathrm{b}\mathrm{b}$ or not because there are still two possibilities, $\pmb{x} = \mathrm{adbb}\ldots$ and $\pmb{x} = \mathrm{adba}\ldots$.
Also, in this case, the decoder can distinguish them by reading the following two bits, that is, the decoder can uniquely determine $x_1x_2x_3x_4 \neq \mathrm{a}\mathrm{d}\mathrm{b}\mathrm{b}$ immediately after reading $f_0(\mathrm{a})f_0(\mathrm{d})f_1(\mathrm{b})f_1(\mathrm{b})00 = 1001111100$ since a $00$ cannot follow after encoding $\mathrm{b}$ with $f_1$ in the AIFV code $F^{(\mathrm{II})}$.
In general, for the AIFV code $F^{(\mathrm{II})}$, the following condition holds for any source sequence $\pmb{x} = x_1x_2\ldots x_n$:
at the time immediately after the decoder reads the prefix $b_1b_2\ldots b_k$ of the codeword sequence $f(\pmb{x})$, if $b_1b_2\ldots b_k = f(x'_1)f(x'_2) \ldots f(x'_l)$, 
then the decoder can uniquely determine whether $x'_1x'_2\ldots x'_l$ is a prefix of the original source sequence $\pmb{x}$ or not by reading the following $2$ bits.
Namely, $F^{(\mathrm{II})}$ is a \emph{$2$-bit delay decodable code} (defined formally in Subsection \ref{subsec:k-bitdelay}).

The average codeword length of an AIFV code $F(f_0, f_1, \trans_0, \trans_1)$ is calculated as follows.
Let $I_i \in \{0, 1\}$ be the index of the code table used to encode the $i$-th symbol of the source sequence, for $i = 1, 2, 3, \ldots$.
Then $I_i, i = 1, 2, 3, \ldots$ is a Markov process.
Let $(\pi_0, \pi_1)$ be the stationary probability of the Markov process, that is,
\begin{equation}
\pi_0 = \frac{Q_{1, 0}}{Q_{0, 1} + Q_{1, 0}}, \quad \pi_1 = \frac{Q_{0, 1}}{Q_{0, 1} + Q_{1, 0}}, 
\end{equation}
where $Q_{i, j}$ is the probability of using $f_j$ immediately after using $f_i$, for $i, j \in \{0, 1\}$.
Then the average codeword length $L(F)$ of an AIFV code $F$ is
$L(F) = \pi_0 L_0(F) + \pi_1 L_1(F)$,
where $L_0(F)$ (resp.\ $L_1(F)$) is the average codeword length of the single code table $f_0$ (resp.\ $f_1$).
Thus, the average codeword length of the AIFV code $F^{(\mathrm{II})}$ in Table \ref{tab:code-aifv} is
\begin{eqnarray}
L(F^{(\mathrm{II})}) = \frac{Q_{1, 0}}{Q_{0, 1} + Q_{1, 0}} L_0(F) + \frac{Q_{0, 1}}{Q_{0, 1} + Q_{1, 0}} L_1(F) 
= 1.86666\ldots,
 \end{eqnarray}
which is shorter than $L(f^{(I)})$ of the Huffman code $f^{(\mathrm{I})}$ in Table \ref{tab:code-aifv} (cf.  (\ref{eq:rrg7p2o0qoyd})).
We \cite{Hashimoto2021} indicate that binary AIFV code achieves the optimal average codeword length in the class of $2$-bit delay decodable code with two code tables.

\begin{table}
\caption{An example of an AIFV code $F^{(\mathrm{II})}(f_0, f_1, \trans_0, \trans_1)$.}
\label{tab:code-aifv}
\centering
\begin{tabular}{clclclc}
\hline
$s \in \mathcal{S}$ & $f_0(s)$ & $\trans_0(s)$ & $f_1(s)$ & $\trans_1(s)$\\
\hline \hline
a & 100 & 0 & 1100 & 0\\
b & 00 & 0 & 11 & 1\\
c & 01 & 0 & 01 & 0\\
d & 1 & 1 & 10 & 0\\
\hline
\end{tabular}
\end{table}

Binary AIFV code is generalized to binary AIFV-$m$ code, which can achieve a shorter average codeword length than binary AIFV code for $m \geq 3$, allowing $m$ code tables and at most $m$-bit decoding delay \cite{Hu2017}.
Analyses of the worst-case redundancy of AIFV and AIFV-$m$ codes are studied in the literature\cite{Hu2017, Fujita2020} for $m=2,3,4,5$. 
The papers \cite{IY:ISITA16, IY:ITW17, Fujita2019, Fujita2018, ISIT2018, ISITA2018, Golin2019, Golin2020, ISIT2020, Golin2021, Golin2022, Sumigawa2017, Hashimoto2019, ITW2020} propose the code construction and coding method of AIFV and AIFV-$m$ codes.
Extensions of AIFV-$m$ codes are proposed in the papers \cite{Sugiura2018, Sugiura2022}.

As stated above, there exist $m$-bit delay decodable codes with a finite number of code tables better than Huffman code for $m \geq 2$. However, it is not known for the case $m = 1$.
In this paper, we prove that there are no $1$-bit delay decodable codes with a finite number of code tables, which achieves a shorter average codeword length than the Huffman code.
Namely, Huffman code is optimal in the class of $1$-bit delay decodable codes with a finite number of code tables.

\section{Preliminaries}
\label{sec:preliminary}

First, we define some notations as follows.
Let $|\mathcal{A}|$ denote the cardinality of a finite set $\mathcal{A}$.
Let $\mathcal{A} \times \mathcal{B}$ denote the set of all ordered pairs $(a, b)$,
$a \in \mathcal{A}$ and $b \in \mathcal{B}$, 
that is, $\mathcal{A} \times \mathcal{B} := \{(a, b) : a \in \mathcal{A}, b \in \mathcal{B}\}$.
Let $\mathcal{A}^k$ denote the set of all sequences of length $k$ over a set $\mathcal{A}$,
and let $\mathcal{A}^{\ast}$ denote the set of all sequences of finite length over a set $\mathcal{A}$,
that is, $\mathcal{A}^{\ast} := \mathcal{A}^0 \cup \mathcal{A}^1 \cup \mathcal{A}^2 \cup \cdots$.
The empty sequence is denoted by $\lambda$.
The length of a sequence $\pmb{x}$ is denoted by $|\pmb{x}|$.
In particular, $|\lambda| = 0$.
We say $\pmb{x} \preceq \pmb{y}$ if $\pmb{x}$ is a prefix of $\pmb{y}$, that is, there exists a sequence $\pmb{z}$, possibly $\pmb{z} = \lambda$, such that $\pmb{y} = \pmb{x}\pmb{z}$.
For a non-empty sequence $\pmb{x} = x_1x_2\ldots x_n$, we define $\suff(\pmb{x}) := x_2x_3\ldots x_n$, that is, $\suff(\pmb{x})$ is the sequence obtained by deleting the first letter $x_1$ from $\pmb{x}$. 
Some notations used in this paper are listed in Appendix B.

\subsection{Code-Tuple}
\label{subsec:treepair}

We formalize a time-variant encoder with a finite number of code tables as a \emph{code-tuple}.
An $m$-code-tuple consists of $m$ code tables $f_0, f_1 \ldots, f_{m-1}$ from $\mathcal{S}$ to $\mathcal{C}^{\ast}$ and $m$ mappings $\trans_0, \trans_1, \ldots, \trans_{m-1}$ from $\mathcal{S}$ to $[m] := \{0, 1, 2, \ldots, m-1\}$.
The $m$ mappings $\trans_0, \trans_1, \ldots, \trans_{m-1}$ determine which code table to use to encode the $i$-th symbol $x_i$ of a source sequence:
if the previous symbol $x_{i-1}$ is encoded with $f_j$, then the current symbol $x_i$ is encoded with $f_{\trans_j(x_{i-1})}$.

\begin{definition}
  \label{def:treepair}
Let $m$ be a positive integer.
An \emph{$m$-code-tuple} $F(f_i, \trans_i : i \in [m])$ is a tuple of
$m$ mappings $f_i : \mathcal{S} \rightarrow \mathcal{C}^{\ast}, i \in [m]$ and $m$ mappings $\trans_i : \mathcal{S} \rightarrow [m], i \in [m]$.
\end{definition}

Let $\mathscr{F}^{(m)}$ denote the set of all $m$-code-tuples and define $\mathscr{F} :=  \mathscr{F}^{(1)} \cup \mathscr{F}^{(2)} \cup \mathscr{F}^{(3)} \cup \cdots$.
An element of $\mathscr{F}$ is called a \emph{code-tuple}.
We sometimes write $F(f_i, \trans_i : i \in [m])$ as $F(f, \trans)$ or $F$ for simplicity.
For $F \in \mathscr{F}^{(m)}$, let $|F|$ denote the number of code tables of $F$, that is, $|F| := m$.
We write $[|F|]$ as $[F]$ for simplicity.

\begin{table}
\caption{Four examples of a code-tuple $F^{(\alpha)}(f^{(\alpha)}, \trans^{(\alpha)}), F^{(\beta)}(f^{(\beta)}, \trans^{(\beta)}), F^{(\gamma)}(f^{(\gamma)}, \trans^{(\gamma)})$, and $F^{(\delta)}(f^{(\delta)}, \trans^{(\delta)})$.}
\label{tab:code-tuple}
\centering

\begin{tabular}{clclclclc}
\hline
$s \in \mathcal{S}$ & $f^{(\alpha)}_0(s)$ & $\trans^{(\alpha)}_0(s)$ & $f^{(\alpha)}_1(s)$ & $\trans^{(\alpha)}_1(s)$ & $f^{(\alpha)}_2(s)$ & $\trans^{(\alpha)}_2(s)$ & $f^{(\alpha)}_3(s)$ & $\trans^{(\alpha)}_3(s)$\\
\hline \hline
a & $\lambda$ & 1 & 110 & 3 & 010 & 0 & $\lambda$ & 3\\
b & 000 & 1 & $\lambda$ & 2 & 011 & 1 & $\lambda$ & 3\\
c & 001 & 2 & 110 & 1 & 10 & 2 & $\lambda$ & 3\\
\hline
\end{tabular}
\vspace{8pt}

\begin{tabular}{clclclc}
\hline
$s \in \mathcal{S}$ & $f^{(\beta)}_0(s)$ & $\trans^{(\beta)}_0(s)$ & $f^{(\beta)}_1(s)$ & $\trans^{(\beta)}_1(s)$ & $f^{(\beta)}_2(s)$ & $\trans^{(\beta)}_2(s)$\\
\hline \hline
a & 11 & 2 & 0110 & 1 & 10 & 2\\
b & $\lambda$ & 1 & 0110 & 2 & 11 & 1\\
c & 101 & 1 & 01 & 1 & 1000 & 0\\
d & 1011 & 1 & 0111 & 2 & 1001 & 1\\
e & 1101 & 2 & 01110 & 2 & 11100 & 2\\
\hline
\end{tabular}
\vspace{8pt}

\begin{tabular}{clclclc}
\hline
$s \in \mathcal{S}$ & $f^{(\gamma)}_0(s)$ & $\trans^{(\gamma)}_0(s)$ & $f^{(\gamma)}_1(s)$ & $\trans^{(\gamma)}_1(s)$ & $f^{(\gamma)}_2(s)$ & $\trans^{(\gamma)}_2(s)$\\
\hline \hline
a & 111 & 2 & 1100 & 1 & 01 & 2\\
b & 0 & 1 & 1101 & 2 & 10 & 1\\
c & 1010 & 1 & 10 & 1 & 000 & 0\\
d & 10110 & 1 & 1111 & 2 & 0010 & 1\\
e & 11011 & 2 & 11101 & 2 & 11001 & 2\\
\hline
\end{tabular}
\vspace{8pt}

\begin{tabular}{clclclc}
\hline
$s \in \mathcal{S}$ & $f^{(\delta)}_0(s)$ & $\trans^{(\delta)}_0(s)$ & $f^{(\delta)}_1(s)$ & $\trans^{(\delta)}_1(s)$ & $f^{(\delta)}_2(s)$ & $\trans^{(\delta)}_2(s)$\\
\hline \hline
a & 111 & 2 & 1001 & 1 & 01 & 2\\
b & 01 & 1 & 101 & 2 & 101 & 1\\
c & 10101 & 1 & 01 & 1 & 000 & 0\\
d & 101101 & 1 & 111 & 2 & 00101 & 1\\
e & 11011 & 2 & 1101 & 2 & 11001 & 2\\
\hline
\end{tabular}
\end{table}

\begin{example}
Table \ref{tab:code-tuple} shows four examples of a code-tuple $F^{(\alpha)}(f^{(\alpha)}, \trans^{(\alpha)}) \in \mathscr{F}^{(4)}$ for $\mathcal{S} = \{\mathrm{a}, \mathrm{b}, \mathrm{c}\}$, and $F^{(\beta)}(f^{(\beta)}, \trans^{(\beta)}), F^{(\gamma)}(f^{(\gamma)}, \trans^{(\gamma)}), F^{(\delta)}(f^{(\delta)}, \trans^{(\delta)}) \in \mathscr{F}^{(3)}$ for $\mathcal{S} = \{\mathrm{a}, \mathrm{b}, \mathrm{c}, \mathrm{d}, \mathrm{e}\}$.
\end{example}

In encoding $x_1x_2 \ldots x_n \in \mathcal{S}^{\ast}$ with a $F(f, \trans) \in \mathscr{F}$, the mappings $\trans_0, \trans_1, \ldots, \trans_{|F|-1}$ determine which code table to use to encode $x_i$ for $i = 2, 3, \ldots, n$.
However, there are choices of which code table to use for the first symbol $x_1$.
For $i \in [F]$ and $\pmb{x} \in \mathcal{S}^{\ast}$, we define $f^{\ast}_i(\pmb{x}) \in \mathcal{C}^{\ast}$ as the codeword sequence in the case where $x_1$ is encoded with $f_i$.
Namely, $f^{\ast}_i(x_1x_2x_3\ldots) = f_i(x_1)f_{\trans_i(x_1)}(x_2)f_{\trans_{\trans_i(x_1)}(x_2)}(x_3)\ldots$.
Also, we define $\trans^{\ast}_i(\pmb{x}) \in [F]$ as the index of the code table to be used next after encoding $\pmb{x}$ in the case where $x_1$ is encoded with $f_i$.
We give formal definitions of $f^{\ast}_i$ and $\trans^{\ast}_i$ in the following Definition \ref{def:f_T} as recursive formulas.

\begin{definition}
 \label{def:f_T}
For $F(f, \trans) \in \mathscr{F}$ and $i \in [F]$, we define a mapping $f_i^{\ast} : \mathcal{S}^{\ast} \rightarrow \mathcal{C}^{\ast}$ and a mapping $\trans_i^{\ast} : \mathcal{S}^{\ast} \rightarrow [F]$ as follows.

\begin{equation}
\label{eq:fstar}
f_i^{\ast}(\pmb{x}) = 
\begin{cases}
\lambda &\,\,\text{if}\,\, \pmb{x} = \lambda,\\
f_i(x_1)f_{\trans_i(x_1)}^{\ast}(\suff(\pmb{x})) &\,\,\text{if}\,\, \pmb{x} \neq \lambda,\\
\end{cases}
\end{equation}

\begin{equation}
\label{eq:tstar}
\trans_i^{\ast}(\pmb{x}) = 
\begin{cases}
i &\,\,\text{if}\,\, \pmb{x} = \lambda,\\
\trans^{\ast}_{\trans_i(x_1)}(\suff(\pmb{x})) &\,\,\text{if}\,\, \pmb{x} \neq \lambda,\\
\end{cases}
\end{equation}
for $\pmb{x} = x_1 x_2 \ldots x_n \in \mathcal{S}^{\ast}$.
\end{definition}

\begin{example}
Let $F(f, \trans)$ be $F^{(\beta)}(f^{(\beta)}, \trans^{(\beta)})$ of Table \ref{tab:code-tuple}. We have
\begin{eqnarray}
f_2^{\ast}(\mathrm{baed}) &=& f_2(\mathrm{b})f_1^{\ast}(\mathrm{aed})\\
&=& f_2(\mathrm{b})f_1(\mathrm{a})f_1^{\ast}(\mathrm{ed})\\
&=& f_2(\mathrm{b})f_1(\mathrm{a})f_1(\mathrm{e})f_2^{\ast}(\mathrm{d})\\
&=& f_2(\mathrm{b})f_1(\mathrm{a})f_1(\mathrm{e})f_2(\mathrm{d})f_1^{\ast}(\lambda)\\
&=& f_2(\mathrm{b})f_1(\mathrm{a})f_1(\mathrm{e})f_2(\mathrm{d})\\
&=& 110110011101001.
\end{eqnarray}
\begin{eqnarray}
\trans_2^{\ast}(\mathrm{baed}) = \trans^{\ast}_{1}(\mathrm{aed})
= \trans^{\ast}_{1}(\mathrm{ed})
= \trans^{\ast}_{2}(\mathrm{d})
= \trans^{\ast}_{1}(\mathrm{\lambda})
= 1.
\end{eqnarray}
\end{example}

Directly from Definition \ref{def:f_T}, we obtain the following lemma.
\begin{lemma}
\label{lem:f_T}
For any $F(f, \trans) \in \mathscr{F}$ and $i \in [F]$, the following conditions (i)--(iii) hold.
\begin{enumerate}[(i)]
\item For any $\pmb{x}, \pmb{y} \in \mathcal{S}^{\ast}$, 
$f_i^{\ast}(\pmb{x}\pmb{y}) = f_i^{\ast}(\pmb{x})f^{\ast}_{\trans_i^{\ast}(\pmb{x})}(\pmb{y})$.
\item For any $\pmb{x}, \pmb{y} \in \mathcal{S}^{\ast}$, $\trans_i^{\ast}(\pmb{x}\pmb{y}) = \trans^{\ast}_{\trans^{\ast}_i(\pmb{x})}(\pmb{y})$.
\item For any $\pmb{x}, \pmb{y} \in \mathcal{S}^{\ast}$, 
if $\pmb{x} \preceq \pmb{y}$, then $f^{\ast}_i(\pmb{x}) \preceq f^{\ast}_i(\pmb{y})$.
\end{enumerate}
\end{lemma}

For the code-tuple $F^{(\alpha)}$ in Table \ref{tab:code-tuple}, we can see that ${{f^{(\alpha)}_3}^\ast}(\pmb{x}) = \lambda$ for any $\pmb{x} \in \mathcal{S}^{\ast}$.
To distinguish such abnormal and useless code-tuples from the others, we introduce a class $\mathscr{F}_{\ext}$ in the following Definition \ref{def:F_ext}.\begin{definition}
\label{def:F_ext}
We define $\mathscr{F}_{\ext}$ as the set of all $F(f, \trans) \in \mathscr{F}$ such that 
for any $i \in [F]$, there exists $\pmb{x} \in \mathcal{S}^{\ast}$ such that $|f^{\ast}_i(\pmb{x})| > 0$.
\end{definition}
Directly from Definition \ref{def:F_ext},
for any $F(f, \trans) \in \mathscr{F}_{\ext}, i \in [F]$ and an integer $k \geq 0$,
there exists $\pmb{x} \in \mathcal{S}^{\ast}$ such that $|f^{\ast}_i(\pmb{x})| \geq k$.
Namely, we can extend $f^{\ast}_i(\pmb{x})$ as long as we want by appending symbols to $\pmb{x}$ appropriately. 
The subscription "$\ext$" of $\mathscr{T}_{\ext}$ is an abbreviation of ``extendable.''

\begin{example}
Consider $F^{(\alpha)}, F^{(\beta)}, F^{(\gamma)}$, and $F^{(\delta)}$ of Table \ref{tab:code-tuple}.
We have $F^{(\alpha)} \not \in \mathscr{F}_{\ext}$ because $|{{f^{(\alpha)}_3}^\ast}(\pmb{x})| = 0$ for any $\pmb{x} \in \mathcal{S}^{\ast}$.
On the other hand, $F^{(\beta)}, F^{(\gamma)}, F^{(\delta)} \in \mathscr{F}_{\ext}$.
\end{example}

\subsection{$k$-bit Delay Decodable Code-Tuple}
\label{subsec:k-bitdelay}

Let $F(f, \trans) \in \mathscr{F}$ and $i \in [F]$.
Consider a situation that a source sequence $\pmb{x'} \in \mathcal{S}^{\ast}$ is encoded with $F$ starting from the code table $f_i$.
Then the source sequence $\pmb{x'}$ is encoded to the codeword sequence $f^{\ast}_i(\pmb{x'})$, and the decoder reads it bit by bit from the beginning.
Let $\pmb{b} \preceq f(\pmb{x'})$ be the sequence the decoder has read by a certain moment of the decoding process.
If $\pmb{b} = f(\pmb{x})$ for some $\pmb{x} \in \mathcal{S}^{\ast}$, then there are two possible cases, $\pmb{x} \preceq \pmb{x'}$ and $\pmb{x} \not\preceq \pmb{x'}$.
We say that $F$ is \emph{$k$-bit delay decodable} if it is always possible for the decoder to distinguish the two cases, $\pmb{x} \preceq \pmb{x'}$ and $\pmb{x} \not\preceq \pmb{x'}$, by reading the following $k$ bits $\pmb{c} \in \mathcal{C}^k$ of the codeword sequence $f(\pmb{x})$,
that is, for any pair $(\pmb{x}, \pmb{c}) \in \mathcal{S}^{\ast} \times \mathcal{C}^k$, the decoder can distinguish the two cases,  $\pmb{x} \preceq \pmb{x'}$ and $\pmb{x} \not\preceq \pmb{x'}$ according to the pair $(\pmb{x}, \pmb{c})$.
Thus, $F$ is $k$-bit delay decodable if and only if for any pair $(\pmb{x}, \pmb{c}) \in \mathcal{S}^{\ast} \times \mathcal{C}^k$, it holds that $(\pmb{x}, \pmb{c})$ is \emph{$f^{\ast}_i$-positive} or \emph
{$f^{\ast}_i$-negative} defined as follows.
\begin{definition}
\label{def:pos-neg}
Let $F(f, \trans) \in \mathscr{F}$ and $i \in [F]$.
\begin{enumerate}[(i)]
\item A pair $(\pmb{x}, \pmb{c}) \in \mathcal{S}^{\ast} \times \mathcal{C}^{\ast}$ is said to be \emph{$f^{\ast}_i$-positive} if
for any $\pmb{x'} \in \mathcal{S}^{\ast}$, if $f^{\ast}_i(\pmb{x})\pmb{c} \preceq f^{\ast}_i(\pmb{x'})$, 
then $\pmb{x} \preceq \pmb{x'}$.
\item A pair $(\pmb{x}, \pmb{c}) \in \mathcal{S}^{\ast} \times \mathcal{C}^{\ast}$ is said to be \emph{$f^{\ast}_i$-negative} if
for any $\pmb{x'} \in \mathcal{S}^{\ast}$, if $f^{\ast}_i(\pmb{x})\pmb{c} \preceq f^{\ast}_i(\pmb{x'})$, 
then $\pmb{x} \not\preceq \pmb{x'}$.
\end{enumerate}
\end{definition}

 \begin{definition}
  \label{def:k-bitdelay}
 Let $F(f, \trans) \in \mathscr{F}$ and let $k \geq 0$ be an integer. The code-tuple $F$ is said to be \emph{$k$-bit delay decodable} if
 for any $i \in [F]$ and $(\pmb{x}, \pmb{c}) \in \mathcal{S}^{\ast} \times \mathcal{C}^k$, the pair $(\pmb{x}, \pmb{c})$ is $f^{\ast}_i$-positive or $f^{\ast}_i$-negative.
 For an integer $k \geq 0$, we define $\mathscr{F}_{k\hdec}$ as the set of all $k$-bit delay decodable code-tuples.
 \end{definition}

Note that it is possible that a $(\pmb{x}, \pmb{c}) \in \mathcal{S}^{\ast} \times \mathcal{C}^{\ast}$ is $f^{\ast}_i$-positive and $f^{\ast}_i$-negative simultaneously. 
A $(\pmb{x}, \pmb{c}) \in \mathcal{S}^{\ast} \times \mathcal{C}^{\ast}$ is $f^{\ast}_i$-positive and $f^{\ast}_i$-negative simultaneously if and only if there is no sequence $\pmb{x'}$ satisfying $f^{\ast}_i(\pmb{x})\pmb{c} \preceq f^{\ast}_i(\pmb{x'})$.

In fact, the classes $\mathscr{F}_{k\hdec}, k = 0, 1, 2, \ldots$ form a hierarchical structure 
$\mathscr{F}_{0\hdec} \subseteq \mathscr{F}_{1\hdec} \subseteq \mathscr{F}_{2\hdec} \subseteq \cdots.$
Namely, the following Lemma \ref{lem:hierarchy} holds.
\begin{lemma}
\label{lem:hierarchy}
For any two non-negative integers $k, k'$ such that $k \leq k'$, we have $\mathscr{F}_{k\hdec} \subseteq \mathscr{F}_{k'\hdec}$.
\end{lemma}
\emph{Proof of Lemma \ref{lem:hierarchy}}:
Let $F(f, \trans) \in \mathscr{F}_{k\hdec}$. Fix $i \in [F]$ and $(\pmb{x}, \pmb{c'}) \in \mathcal{S}^{\ast} \times \mathcal{C}^{k'}$ arbitrarily. 
It suffices to prove that $(\pmb{x}, \pmb{c'})$ is $f^{\ast}_i$-positive or $f^{\ast}_i$-negative.

Let $\pmb{c}$ be the prefix of $\pmb{c'}$ of length $k$.
Then for any $\pmb{x'} \in \mathcal{S}^{\ast}$ such that $f^{\ast}_i(\pmb{x})\pmb{c'} \preceq f^{\ast}_i(\pmb{x'})$, we have
$f^{\ast}_i(\pmb{x})\pmb{c} \preceq f^{\ast}_i(\pmb{x})\pmb{c'} \preceq f^{\ast}_i(\pmb{x'}$). Namely, $f^{\ast}_i(\pmb{x})\pmb{c'} \preceq f^{\ast}_i(\pmb{x'})$ implies $f^{\ast}_i(\pmb{x})\pmb{c} \preceq f^{\ast}_i(\pmb{x'})$.
Hence, from Definition \ref{def:pos-neg}, if $(\pmb{x}, \pmb{c'})$ is $f^{\ast}_i$-positive (resp.\ $f^{\ast}_i$-negative), then also $(\pmb{x}, \pmb{c})$ is $f^{\ast}_i$-positive (resp.\ $f^{\ast}_i$-negative), respectively.
Therefore, it follows that $F(f, \trans) \in \mathscr{F}_{k'\hdec}$ from $F(f, \trans) \in \mathscr{F}_{k\hdec}$. \qed

\begin{example}
Consider $F^{(\alpha)}, F^{(\beta)}, F^{(\gamma)}$, and $F^{(\delta)}$ of Table \ref{tab:code-tuple}.
We have $F^{(\alpha)} \in \mathscr{F}_{2\hdec} \setminus \mathscr{F}_{1\hdec}$, $F^{(\beta)} \in \mathscr{F}_{1\hdec} \setminus \mathscr{F}_{0\hdec}$, and $F^{(\gamma)}, F^{(\delta)} \in \mathscr{F}_{0\hdec}$.
\end{example}

We remark that a $k$-bit delay decodable code-tuple is not necessarily uniquely decodable.
For example, for the code-tuple $F^{(\beta)} \in \mathscr{F}_{1\hdec}$, we have $f^{\ast}_0(\mathrm{ac}) = 111000 = f^{\ast}_0(\mathrm{acb})$.
In general, it is possible that the decoder cannot uniquely recover the last few symbols of the original source sequence because the length of the rest of the codeword sequence is less than $k$ bits.
In such a case, we should append additional information to uniquely decode the suffix in practical use.

However,  as we show in the following Lemma \ref{lem:0dec-unique}, a $0$-bit delay decodable code-tuple (i.e., an instantaneous code) is always uniquely decodable.
\begin{lemma}
\label{lem:0dec-unique}
For any $F(f, \trans) \in \mathscr{F}_{0\hdec}$ and $i \in [F]$, the following conditions (i) and (ii) hold.
\begin{enumerate}[(i)]
\item For any $\pmb{x} \in \mathcal{S}^{\ast}$, the pair $(\pmb{x}, \lambda)$ is $f^{\ast}_i$-positive.
\item $f^{\ast}_i$ is injective.
\end{enumerate}
\end{lemma}
 \emph{Proof of Lemma \ref{lem:0dec-unique}}:
 (Proof of (i)) From $F \in \mathscr{F}_{0\hdec}$, the pair $(\pmb{x}, \lambda)$ is $f^{\ast}_i$-positive or $f^{\ast}_i$-negative.
 However, since $f^{\ast}_i(\pmb{x}) \preceq f^{\ast}_i(\pmb{x})$ and $\pmb{x} \preceq \pmb{x}$, 
 the pair $(\pmb{x}, \lambda)$ must be $f^{\ast}_i$-positive.
 
(Proof of (ii)) From (i), we have
\begin{equation}
 \label{eq:0dec-unique-2}
{}^\forall \pmb{x}, \pmb{x'} \in \mathcal{S}^{\ast}, 
\big( f^{\ast}_i(\pmb{x}) \preceq f^{\ast}_i(\pmb{x'}) \Rightarrow \pmb{x} \preceq \pmb{x'} \big).
\end{equation}
Choose $\pmb{y}, \pmb{y'} \in \mathcal{S}^{\ast}$ such $f^{\ast}_i(\pmb{y}) = f^{\ast}_i(\pmb{y'})$ arbitrarily.
Then we have $f^{\ast}_i(\pmb{y}) \preceq f^{\ast}_i(\pmb{y'})$ and $f^{\ast}_i(\pmb{y'}) \preceq f^{\ast}_i(\pmb{y})$.
From (\ref{eq:0dec-unique-2}), we obtain $\pmb{y} \preceq \pmb{y'}$ and $\pmb{y'} \preceq \pmb{y}$, that is, $\pmb{y} = \pmb{y'}$.
Consequently, $f^{\ast}_i$ is injective.
\qed.

For $F(f, \trans) \in \mathscr{F}$ and $i \in [F]$, $f_i$ is said to be \emph{prefix-free} if
for any $s, s' \in \mathcal{S}$, if $f_i(s) \preceq f_i(s')$, then $s = s'$.
A $0$-bit delay decodable code-tuple is characterized as a code-tuple all of which code tables are prefix-free.

\begin{lemma}
\label{lem:0dec-prefixfree}
A $F(f, \trans) \in \mathscr{F}$ satisfies $F \in \mathscr{F}_{0\hdec}$ if and only if for any $i \in [F]$, $f_i$ is prefix-free.
\end{lemma}
 \emph{Proof of Lemma \ref{lem:0dec-prefixfree}}:
 (Proof of ``only if'')
 Assume $F \in \mathscr{F}_{0\hdec}$ and choose $i \in [F]$ arbitrarily.
 From Lemma \ref{lem:0dec-unique} (i), the pair $(\pmb{x}, \lambda)$ is $f^{\ast}_i$-positive. Thus, (\ref{eq:0dec-unique-2}) holds.
 In particular, we have
\begin{equation}
{}^\forall s, s' \in \mathcal{S}, 
\big( f^{\ast}_i(s) \preceq f^{\ast}_i(s') \Rightarrow s \preceq s' \big).
\end{equation}
Since $s \preceq s'$ implies $s = s'$, $f_i$ is prefix-free. 

 (Proof of ``if'')
 Assume that for any $i \in [F]$, $f_i$ is prefix-free. Fix $i \in [F]$ arbitrarily.
To prove $F \in \mathscr{F}_{0\hdec}$, it suffices to prove (\ref{eq:0dec-unique-2}). We prove it by induction for $|\pmb{x}|$.

For the case $|\pmb{x}| = 0$, clearly we have $\pmb{x} \preceq \pmb{x'}$ for any $\pmb{x'} \in \mathcal{S}^{\ast}$.

Let $l \geq 1$ and assume that (\ref{eq:0dec-unique-2}) is true for the case $|\pmb{x}| < l$ as the induction hypothesis. We prove (\ref{eq:0dec-unique-2}) for the case $|\pmb{x}| = l$.
Choose $\pmb{x'} \in \mathcal{S}^{\ast}$ such that $f^{\ast}_i(\pmb{x}) \preceq f^{\ast}_i(\pmb{x'})$ arbitrarily.
Then by (\ref{eq:fstar}), we have
\begin{equation}
\label{eq:0dec-prefixfree}
f_i(x_1)f^{\ast}_{\trans^{\ast}_i(x_1)}(\suff(\pmb{x})) \preceq f_i(x'_1)f^{\ast}_{\trans^{\ast}_i(x'_1)}(\suff(\pmb{x'})),
\end{equation}
where $\pmb{x} = x_1x_2\ldots x_n$, $\pmb{x'} = x'_1x'_2\ldots x'_{n'}$.
Thus, $f_i(x_1) \preceq f_i(x'_1)$ or $f_i(x_1) \succeq f_i(x'_1)$ holds.
Hence, since $f_i$ is prefix-free, we obtain
\begin{equation}
\label{eq:i2x624m8pz92}
x_1 = x'_1.
\end{equation}

From (\ref{eq:0dec-prefixfree}) and (\ref{eq:i2x624m8pz92}), we have
$f_i(x_1)f^{\ast}_{\trans^{\ast}_i(x_1)}(\suff(\pmb{x})) \preceq f_i(x_1)f^{\ast}_{\trans^{\ast}_i(x'_1)}(\suff(\pmb{x'})).$
Thus, we have
$f^{\ast}_{\trans^{\ast}_i(x_1)}(\suff(\pmb{x})) \preceq f^{\ast}_{\trans^{\ast}_i(x'_1)}(\suff(\pmb{x'})).$
From the induction hypothesis,
\begin{equation}
\label{eq:0dec-prefixfree2}
\suff(\pmb{x}) \preceq \suff(\pmb{x'}).
\end{equation}
By (\ref{eq:i2x624m8pz92}) and (\ref{eq:0dec-prefixfree2}), we obtain $\pmb{x} \preceq \pmb{x'}$.
\qed

\subsection{Average Codeword Length of Code-Tuple}
\label{subsec:evaluation}

We introduce the average codeword length $L(F)$ of code-tuple $F$.
Hereinafter, we fix a probability distribution $\mu$ of source symbols, that is, a real-valued function $\mu : \mathcal{S} \rightarrow \mathbb{R}$ 
satisfying $\sum_{s \in \mathcal{S}} \mu(s) = 1$ and $0 < \mu(s) \leq 1$ for any $s \in \mathcal{S}$.
Note that we exclude the case where $\mu(s) = 0$ for some $s \in \mathcal{S}$ from our consideration without loss of generality.

First, we define the transition probability $Q_{i, j}(F)$ for $F(f, \trans) \in \mathscr{F}$ and $i, j \in [F]$ as the probability of using $f_j$ immediately after using $f_i$.

 \begin{definition}
\label{def:transprobability}
For $F(f, \trans) \in \mathscr{F}$ and $i, j \in [F]$,
we define the \emph{transition probability} $Q_{i,j}(F)$ as $Q_{i,j}(F) := \sum_{s \in \mathcal{S}, \trans_i(s) = j} \mu(s)$.
We also define the \emph{transition probability matrix} $Q(F)$ as the following $|F|\times|F|$ matrix.
 \begin{equation}
  Q(F) := \left[
    \begin{array}{cccc}
      Q_{0,0}(F) & Q_{0,1}(F) & \cdots & Q_{0, |F|-1}(F) \\
       Q_{1,0}(F) &  Q_{1,1}(F) & \cdots &  Q_{1, |F|-1}(F) \\
      \vdots & \vdots & \ddots & \vdots \\
      Q_{|F|-1, 0}(F) &  Q_{|F|-1, 1}(F) & \cdots &  Q_{|F|-1, |F|-1}(F) 
    \end{array}
  \right].
 \end{equation}
  \end{definition}

Fix a $F \in \mathscr{F}$ and Let $I_i \in \{0, 1\}$ be the index of the code table used to encode the $i$-th symbol of the source sequence in encoding with $F$ for $i = 1, 2, 3, \ldots$.
Then $I_i \in \{0, 1\}, i = 1, 2, 3, \ldots$ is a Markov process with the transition probability matrix $Q(F)$.
We consider a stationary distribution of the Markov process (i.e., the solution of the simultaneous equations (\ref{eq:stationary1}) and (\ref{eq:stationary2})).
The average codeword length $L(F)$ of a code-tuple $F$ depends on a stationary distribution of the Markov process with $Q(F)$.
Hence, to define $L(F)$ uniquely, a stationary distribution of the Markov process with $Q(F)$ must be unique.
Thus, we define the class $\mathscr{F}_{\reg}$ of all code-tuples such that the Markov process has a unique stationary distribution.

\begin{definition}
\label{def:regular}
A $F \in \mathscr{F}$ is said to be \emph{regular} if
the following simultaneous equations (\ref{eq:stationary1}) and (\ref{eq:stationary2}) have a unique solution $\pmb{\pi} = (\pi_0, \pi_1, \ldots, \pi_{|F|-1})$.
\begin{numcases}{}
\pmb{\pi}Q(F) = \pmb{\pi}\label{eq:stationary1},\\
 \sum_{i \in [F]} \pi_i = 1. \label{eq:stationary2}
 \end{numcases}
 
 We define $\mathscr{F}_{\reg}$ as the set of all regular code-tuples.
 For $F \in \mathscr{F}_{\reg}$, we define $\pmb{\pi}(F) = (\pi_0(F), \pi_1(F), \ldots, \pi_{|F|-1}(F))$
 as the unique solution of the simultaneous equations (\ref{eq:stationary1}) and (\ref{eq:stationary2}).
 \end{definition}
 
For any $F \in \mathscr{F}_{\reg}$, the asymptotical performance (average codeword length per symbol) does not depend on which code table we start encoding:
the average codeword length $L(F)$ of a regular code tuple $F \in \mathscr{F}_{\reg}$ is
the weighted sum of average codeword lengths of a single code table $f_0, f_1, \ldots, f_{|F|-1}$ weighted by the stationary distribution $\pmb{\pi}(F)$.
 
 \begin{definition} 
 \label{def:evaluation}
 For $F(f, \trans) \in \mathscr{F}$ and $i \in [F]$, we define the \emph{average codeword length $L_i(F)$ of a single code table} $f_i : \mathcal{S} \rightarrow \mathcal{C}^{\ast}$ as $L_i(F) := \sum_{s \in \mathcal{S}} |f_i(s)| \cdot \mu(s)$.
For $F \in \mathscr{F}_{\reg}$,
we define the \emph{average codeword length $L(F)$ of code-tuple $F$} as $L(F) := \sum_{i \in [F]} \pi_i(F)L_i(F)$.
\end{definition}

 \begin{example}
 \label{ex:evaluation}
Let $F(f, \trans)$ be $F^{(\beta)}(f^{(\beta)}, \trans^{(\beta)})$ of Table \ref{tab:code-tuple} and let $(\mu(\mathrm{a}), \mu(\mathrm{b}), \mu(\mathrm{c}), \mu(\mathrm{d}), \mu(\mathrm{e})) = (0.1, 0.2, 0.2, 0.2, 0.3)$.

We have 
 \begin{equation}
  Q(F) = \left[
    \begin{array}{cccc}
      0 & 0.6 & 0.4 \\
       0 &  0.3 & 0.7\\
      0.2 &  0.4 & 0.4 
    \end{array}
  \right],
\end{equation}

\begin{equation}
L_0(F) = 2.8, \quad L_1(F) = 3.9, \quad  L_2(F) = 3.7.
\end{equation}
Also, we obtain $\pmb{\pi}(F) = (7/68, 26/68, 35/68)$ by solving the simultaneous equations (\ref{eq:stationary1}) and (\ref{eq:stationary2}).
Therefore, the average codeword length $L(F)$ of the code-tuple $F$ is given as
\begin{eqnarray}
L(F) &=& \pi_0(F)L_0(F) + \pi_1(F)L_1(F) + \pi_2(F)L_2(F) \\
&=& \frac{7 \cdot 2.8 + 26 \cdot 3.9 + 35 \cdot 3.7}{68} \\
&\approx& 3.683823.
\end{eqnarray}
 \end{example}

\section{The Optimality of Huffman Code}
\label{sec:optimality}

In this section, we prove the following Theorem \ref{thm:1dec-Huff} as the main result of this paper.

\setcounter{theorem}{0}
\begin{theorem}
\label{thm:1dec-Huff}
For any $F(f, \trans) \in \mathscr{F}_{\reg} \cap \mathscr{F}_{\ext} \cap \mathscr{F}_{1\hdec}$, we have $L(F) \geq L_{\Huff}$, where $L_{\Huff}$ is the average codeword length of the Huffman code.
\end{theorem}

The outline of the proof is as follows.
First, we define an operation called \emph{rotation} which transforms a code-tuple $F \in \mathscr{F}_{\ext}$ into another code-tuple $\widehat{F} \in \mathscr{F}$.
Then we show that any $F \in \mathscr{F}_{\reg} \cap \mathscr{F}_{\ext} \cap \mathscr{F}_{1\hdec}$ can be transformed into some $F' \in \mathscr{F}_{\reg} \cap\mathscr{F}_{\ext} \cap \mathscr{F}_{0\hdec}$ by rotation in a repetitive manner without changing the average codeword length.
Hence, without loss of generality, we can assume that a given code-tuple is in $\mathscr{F}_{\reg} \cap\mathscr{F}_{\ext} \cap \mathscr{F}_{0\hdec}$, in particular, $0$-bit delay decodable.
Then we complete the proof of Theorem \ref{thm:1dec-Huff} using McMillan's theorem in the context of $0$-bit delay decodable code.
This section consists of the following four subsections:
\begin{enumerate}
\item In subsection \ref{subsec:optimality-1}, we introduce \emph{rotation} which transforms a code-tuple $F \in \mathscr{F}_{\ext}$ into another code-tuple $\widehat{F} \in \mathscr{F}$.
\item In subsection \ref{subsec:optimality-2}, we show that for any $F \in \mathscr{F}_{\reg} \cap \mathscr{F}_{\ext} \cap \mathscr{F}_{1\hdec}$, we have $\widehat{F} \in \mathscr{F}_{\reg} \cap \mathscr{F}_{\ext} \cap \mathscr{F}_{1\hdec}$ and $L(\widehat{F}) = L(F)$.
Namely, the rotation preserves ``the key properties'' of a code-tuple.
\item In subsection \ref{subsec:optimality-3}, we prove that for any $F \in \mathscr{F}_{\reg} \cap \mathscr{F}_{\ext} \cap \mathscr{F}_{1\hdec}$,
there exists $F' \in \mathscr{F}_{\reg} \cap\mathscr{F}_{\ext} \cap \mathscr{F}_{0\hdec}$ such that $L(F') = L(F)$.
Namely, any $F \in \mathscr{F}_{\reg} \cap \mathscr{F}_{\ext} \cap \mathscr{F}_{1\hdec}$ can be replaced with some $F' \in \mathscr{F}_{\reg} \cap\mathscr{F}_{\ext} \cap \mathscr{F}_{0\hdec}$.
\item In subsection \ref{subsec:optimality-4}, we complete the proof of Theorem \ref{thm:1dec-Huff} using McMillan's theorem.
\end{enumerate}

\subsection{Rotation}
\label{subsec:optimality-1}

As stated above, the first step in the proof is to define rotation.
To describe the definition of rotation, we introduce the following Definition \ref{def:pref}.
\begin{definition}
\label{def:pref}
For $F(f, \trans) \in \mathscr{F}$ and $i \in [F]$, we define $\PREF_{F, i}$ 
as $\PREF_{F, i} := \{c \in \mathcal{C} : {}^{\exists} \pmb{x} \in \mathcal{S}^{\ast}, f^{\ast}_i(\pmb{x}) \succeq c\}$,
that is, $\PREF_{F, i}$ is the set of all $c \in \mathcal{C}$ which is the first bit of $f^{\ast}_i(\pmb{x})$ for some $\pmb{x} \in \mathcal{S}^{\ast}$.

For $F \in \mathscr{F}_{\ext}$ and $i \in [F]$, we define $d_{F, i}$ as follows.
\begin{equation}
\label{eq:emksr78o7m54}
d_{F, i} =
\begin{cases}
0 &\,\,\text{if}\,\, \PREF_{F, i} = \{0\},\\
1 &\,\,\text{if}\,\, \PREF_{F, i} = \{1\},\\
\lambda &\,\,\text{if}\,\, \PREF_{F, i} = \{0, 1\}.\\
\end{cases}\end{equation}

\end{definition}

Note that for any $F \in \mathscr{F}_{\ext}$ and $i \in [F]$, we have $\PREF_{F, i} \neq \emptyset$.

\begin{table}
\caption{$\PREF_{F, i}, i \in [F]$ and $d_{F, i}, i \in [F]$ for the code-tuples $F^{(\alpha)}, F^{(\beta)}, F^{(\gamma)}$, and $F^{(\delta)}$ in Table \ref{tab:code-tuple}.}
\label{tab:pref}
\centering
\begin{tabular}{ccccccccc}
\hline
$F$ & $\PREF_{F, 0}$ & $d_{F, 0}$ & $\PREF_{F, 1}$ & $d_{F, 1}$ & $\PREF_{F, 2}$ & $d_{F, 2}$ & $\PREF_{F, 3}$ & $d_{F, 3}$\\
\hline \hline
$F^{(\alpha)}$ & $\{0, 1\}$ & - & $\{0, 1\}$ & - & $\{0, 1\}$ & -& $\emptyset$ & -\\
$F^{(\beta)}$ & $\{0, 1\}$ & $\lambda$ & $\{0\}$ & $0$ & $\{1\}$ & $1$ & - & -\\
$F^{(\gamma)}$ & $\{0, 1\}$ & $\lambda$ & $\{1\}$ & $1$ & $\{0, 1\}$ & $\lambda$ & - & -\\
$F^{(\delta)}$ & $\{0, 1\}$ & $\lambda$ & $\{0, 1\}$ & $\lambda$  & $\{0, 1\}$ & $\lambda$ & - & -\\
\hline
\end{tabular}
\end{table}

\begin{example}
\label{ex:pref}
Table \ref{tab:pref} shows $\PREF_{F, i}, i \in [F]$ and $d_{F, i}, i \in [F]$ for the code-tuples $F^{(\alpha)}, F^{(\beta)}, F^{(\gamma)}$, and $F^{(\delta)}$ in Table \ref{tab:code-tuple}.
Note that $d_{F^{(\alpha)}, 0}, d_{F^{(\alpha)}, 1}$ and $d_{F^{(\alpha)}, 2}$ are not defined since $F^{(\alpha)} \not\in \mathscr{F}_{\ext}$.
\end{example}

Then the rotation is defined as follows.

\begin{definition}
\label{def:rotation}
For $F(f, \trans) \in \mathscr{F}_{\ext}$, we define $\widehat{F}(\widehat{f}, \widehat{\trans}) \in \mathscr{F}$ as follows.

For $i \in [F]$ and $s \in \mathcal{S}$,
\begin{equation}
\label{eq:8ckwdpt4s2b5}
\widehat{f}_i(s) = 
\begin{cases}
f_i(s)d_{F, \trans_i(s)} &\,\,\text{if}\,\, \PREF_{F, i} = \{0, 1\},\\
\suff(f_i(s)d_{F, \trans_i(s)}) &\,\,\text{if}\,\, \PREF_{F, i} \neq \{0, 1\},\\
\end{cases}
\end{equation}
\begin{equation}
\label{eq:g8ovoqynvwxb}
\widehat{\trans}_i(s) = \trans_i(s).
\end{equation}
The operation which transforms a given $F \in \mathscr{F}_{\ext}$ into $\widehat{F} \in \mathscr{F}$ defined above is called \emph{rotation}.
\end{definition}

\begin{example}
\label{ex:rotation}
In Table \ref{tab:code-tuple}, $F^{(\gamma)}$ is obtained by applying rotation to $F^{(\beta)}$, that is, $F^{(\gamma)} = \widehat{F^{(\beta)}}$.
Also, $F^{(\delta)}$ is obtained by applying rotation to $F^{(\gamma)}$, that is, $F^{(\delta)} = \widehat{F^{(\gamma)}}$.
Furthermore, we obtain $F^{(\delta)}$ itself applying rotation to $F^{(\delta)}$, that is, $F^{(\delta)} = \widehat{F^{(\delta)}}$.
\end{example}

Directly from Definition \ref{def:rotation}, we can see that for any $F(f, \trans) \in \mathscr{F}_{\ext}, i \in [F]$, and $s \in \mathcal{S}$, we have
\begin{equation}
\label{eq:rotation}
d_{F, i}\widehat{f_i}(s) = f_i(s)d_{F, \trans_i(s)}.
\end{equation}
We show that (\ref{eq:rotation}) is generalized to the following Lemma \ref{lem:rotation}.

\begin{lemma}
\label{lem:rotation}
For any $F(f, \trans) \in \mathscr{F}_{\ext}, i \in [F]$, and $\pmb{x} \in \mathcal{S}^{\ast}$,
\begin{equation}
\label{eq:lem-rotation}
d_{F, i}\widehat{f^{\ast}_i}(\pmb{x}) = f^{\ast}_i(\pmb{x})d_{F, \trans^{\ast}_i(\pmb{x})}.
\end{equation}
\end{lemma}
\emph{Proof of Lemma \ref{lem:rotation}}: We prove the lemma by induction for $|\pmb{x}|$.

For the case $|\pmb{x}| = 0$, we have
\begin{equation}
d_{F, i}\widehat{f^{\ast}_i}(\pmb{x}) = d_{F, i}\widehat{f^{\ast}_i}(\lambda)
= d_{F, i}\lambda = \lambda d_{F, i} = f^{\ast}_i(\lambda)d_{F, \trans^{\ast}_i(\lambda)}
= f^{\ast}_i(\pmb{x})d_{F, \trans^{\ast}_i(\pmb{x})},
\end{equation}
where the second and forth equalities are from (\ref{eq:fstar}).
Hence, (\ref{eq:lem-rotation}) holds.

Let $l \geq 1$ and assume that (\ref{eq:lem-rotation}) holds for any $\pmb{x'} \in \mathcal{S}^{\ast}$ such that $|\pmb{x'}| < l$ as the induction hypothesis.
We prove that (\ref{eq:lem-rotation}) holds for $\pmb{x} := x_1x_2\ldots x_l \in \mathcal{S}^{\ast}$.
We have 
\begin{eqnarray}
\lefteqn{d_{F, i}\widehat{f^{\ast}_i}(\pmb{x})} \nonumber\\
&=& d_{F, i}\widehat{f_i}(x_1)\widehat{f^{\ast}}_{\trans_i(x_1)}(\suff(\pmb{x}))\\
&=& f_i(x_1)d_{F, \trans_i(x_1)}\widehat{f^{\ast}}_{\trans_i(x_1)}(\suff(\pmb{x}))\\
&=& f_i(x_1)f^{\ast}_{\trans_i(x_1)}(\suff(\pmb{x}))d_{F, \trans^{\ast}_{\trans_i(x_1)}(\suff(\pmb{x}))}\\
&=& f_i(x_1)f^{\ast}_{\trans_i(x_1)}(\suff(\pmb{x}))d_{F, \trans^{\ast}_i(\pmb{x})}\\
&=& f^{\ast}_i(\pmb{x})d_{F, \trans^{\ast}_i(\pmb{x})},
\end{eqnarray}
where the first equality is from (\ref{eq:fstar}), the second equality is from (\ref{eq:rotation}), the third equality is from the induction hypothesis,
the forth equality is from (\ref{eq:tstar}) and the fifth equality is from (\ref{eq:fstar}).
Hence, (\ref{eq:lem-rotation}) holds for $\pmb{x} := x_1x_2\ldots x_l \in \mathcal{S}^{\ast}$.
\qed

 \begin{example}
 Let $F(f, \trans)$ be $F^{(\beta)}$ of Table \ref{tab:code-tuple}. 
 As seen in Example \ref{ex:rotation}, we have $\widehat{F}(\widehat{f}, \widehat{\trans}) = F^{(\gamma)}$. 
 We can see
$ d_{F, 2}\widehat{f^{\ast}_2}(\mathrm{baed}) =  d_{F, 2}\widehat{f_2}(\mathrm{b})\widehat{f_1}(\mathrm{a})\widehat{f_1}(\mathrm{e})\widehat{f_2}(\mathrm{d})
 = 1101100111010010,$
 and
$ f^{\ast}_2(\mathrm{baed})d_{F, \trans^{\ast}_2(\mathrm{baed})} = f^{\ast}_2(\mathrm{baed})d_{F, 1} = 1101100111010010.$
Hence, we can confirm $d_{F, 2}\widehat{f^{\ast}_2}(\mathrm{baed}) =  f^{\ast}_2(\mathrm{baed})d_{F, \trans^{\ast}_2(\mathrm{baed})}$.
\end{example}

\subsection{Rotation Preserves the Key Properties}
\label{subsec:optimality-2}

In the previous subsection, we introduced rotation, which transforms a given $F \in \mathscr{F}_{\ext}$ into the $\widehat{F} \in \mathscr{F}$.
In this subsection, we prove that if $F \in \mathscr{F}_{\reg} \cap \mathscr{F}_{\ext} \cap \mathscr{F}_{1\hdec}$, then $\widehat{F} \in \mathscr{F}_{\reg} \cap \mathscr{F}_{\ext} \cap \mathscr{F}_{1\hdec}$ and $L(\widehat{F}) = L(F)$.
To prove it, we show the following Lemmas \ref{lem:rot-decodable}--\ref{lem:rot-equal}.

\begin{lemma}
\label{lem:rot-decodable}
For any integer $k \geq 0$ and $F(f, \trans) \in \mathscr{F}_{k\hdec} \cap \mathscr{F}_{\ext}$, 
we have $\widehat{F}(\widehat{f}, \widehat{\trans}) \in \mathscr{F}_{k\hdec}$.
\end{lemma}
\begin{lemma}
\label{lem:rot-extendable}
For any $F(f, \trans) \in \mathscr{F}_{\ext}$, we have $\widehat{F}(\widehat{f}, \widehat{\trans}) \in \mathscr{F}_{\ext}$.
\end{lemma}
\begin{lemma}
  \label{lem:rot-equal}
For any $F \in \mathscr{F}_{\reg}$, we have $\widehat{F} \in \mathscr{F}_{\reg}$ and $L(\widehat{F}) = L(F).$
\end{lemma}

The proof of Lemma \ref{lem:rot-decodable} relies on Lemma \ref{lem:rot-prefix} whose proof is relegated to Appendix A.

\begin{lemma}
\label{lem:rot-prefix}
Let $F(f, \trans) \in \mathscr{F}_{\ext}$.
There exists no tuple $(k, i, \pmb{x}, \pmb{x'})$ satisfying all of the following conditions (i)--(iii),
where $k$ is a non-negative integer, $i \in [F]$, and $\pmb{x}, \pmb{x'} \in \mathcal{S}^{\ast}$:
(i) $F \in \mathscr{F}_{k\hdec}$, 
(ii) $|\widehat{f^{\ast}_i}(\pmb{x})| + k \leq |\widehat{f^{\ast}_i}(\pmb{x'})|$,
and (iii) $\pmb{x'} \preceq \pmb{x}$ and $\pmb{x} \neq \pmb{x'}$.
\end{lemma}

\noindent
\emph{Proof of Lemma \ref{lem:rot-decodable}}:
Fix $i \in [F]$ and $(\pmb{x}, \pmb{c}) \in \mathcal{S}^{\ast} \times \mathcal{C}^k$ arbitrarily.

Also, choose $\pmb{x'} \in \mathcal{S}^{\ast}$ such that $\widehat{f^{\ast}_i}(\pmb{x})\pmb{c} \preceq \widehat{f^{\ast}_i}(\pmb{x'})$ arbitrarily.
Then, we have
$d_{F, i}\widehat{f^{\ast}_i}(\pmb{x})\pmb{c} \preceq d_{F, i}\widehat{f^{\ast}_i}(\pmb{x'}).$
From Lemma \ref{lem:rotation}, we have
$f^{\ast}_i(\pmb{x})d_{F, \trans^{\ast}_i(\pmb{x})}\pmb{c} \preceq f^{\ast}_i(\pmb{x'})d_{F, \trans^{\ast}_i(\pmb{x'})}.$
From (\ref{eq:emksr78o7m54}), there exists $\pmb{x''} \in \mathcal{S}^{\ast}$ such that $d_{F, \trans^{\ast}_i(\pmb{x'})} \preceq f^{\ast}_{\trans^{\ast}_i(\pmb{x'})}(\pmb{x''})$.
For such $\pmb{x''}$, we have
\begin{equation}
f^{\ast}_i(\pmb{x})d_{F, \trans^{\ast}_i(\pmb{x})}\pmb{c} \preceq f^{\ast}_i(\pmb{x'})d_{F, \trans^{\ast}_i(\pmb{x'})} \preceq f^{\ast}_i(\pmb{x'})f^{\ast}_{\trans^{\ast}_i(\pmb{x'})}(\pmb{x''}) = f^{\ast}_i(\pmb{x'}\pmb{x''}),
\end{equation}
where the last equality is from Lemma \ref{lem:f_T} (i).
Therefore, we obtain
\begin{equation}
\label{eq:rot-decodable}
f^{\ast}_i(\pmb{x})\pmb{c'} \preceq f^{\ast}_i(\pmb{x'}\pmb{x''}),
\end{equation}
where $\pmb{c'}$ is the prefix of length of $k$ of $d_{F, \trans^{\ast}_i(\pmb{x})}\pmb{c}$.

In general, exactly one of the following conditions (a)--(c) holds:
(a) $\pmb{x} \preceq \pmb{x'}$,  (b) $\pmb{x} \neq \pmb{x'}$ and $\pmb{x} \succeq \pmb{x'}$, and (c) $\pmb{x} \not\preceq \pmb{x'}$ and $\pmb{x} \not\succeq \pmb{x'}$.
However, now (b) is impossible from Lemma \ref{lem:rot-prefix}.
Thus, it suffices to consider the case where either (a) or (c) holds.

From $F \in \mathscr{F}_{k\hdec}$, $(\pmb{x}, \pmb{c'})$ is $f^{\ast}_i$-positive or $f^{\ast}_i$-negative.
If $(\pmb{x}, \pmb{c'})$ is $f^{\ast}_i$-positive (resp.\ $f^{\ast}_i$-negative), 
then we have $\pmb{x} \preceq \pmb{x'}\pmb{x''}$ (resp.\ $\pmb{x} \not\preceq \pmb{x'}\pmb{x''}$) from (\ref{eq:rot-decodable}).
This implies that (a) $\pmb{x} \preceq \pmb{x'}$ (resp.\ (c) $\pmb{x} \not\preceq \pmb{x'}$ and $\pmb{x} \not\succeq \pmb{x'}$) holds since (b) is impossible.
Since $\pmb{x'}$ is chosen arbitrarily, the pair $(\pmb{x}, \pmb{c})$ is $\widehat{f^{\ast}_i}$-positive (resp.\ $\widehat{f^{\ast}_i}$-negative), respectively.
Therefore, we have $\widehat{F} \in \mathscr{F}_{k\hdec}$.
\qed

\noindent
\emph{Proof of Lemma \ref{lem:rot-extendable}}:
Fix $i \in [F]$ arbitrarily.
From $F \in \mathscr{F}_{\ext}$, there exists $\pmb{x} \in \mathcal{S}^{\ast}$ such that $|f^{\ast}_i(\pmb{x})| \geq 2$.
For such $\pmb{x}$, from Lemma \ref{lem:rotation},
we have $d_{F, i}\widehat{f^{\ast}_i}(\pmb{x}) = f^{\ast}_i(\pmb{x})d_{F, \trans^{\ast}_i(\pmb{x})}.$
Hence,
$|\widehat{f^{\ast}_i}(\pmb{x})| = |f^{\ast}_i(\pmb{x})| + |d_{F, \trans^{\ast}_i(\pmb{x})}| - |d_{F, i}|.$
From $|f^{\ast}_i(\pmb{x})| \geq 2, |d_{F, \trans^{\ast}_i(\pmb{x})}| \geq 0$, and $|d_{F, i}| \leq 1$, we obtain $|\widehat{f^{\ast}_i}(\pmb{x})| \geq 1.$
Therefore, we have $\widehat{F} \in \mathscr{F}_{\ext}$.
\qed

\noindent
\emph{Proof of Lemma \ref{lem:rot-equal}}:
From (\ref{eq:g8ovoqynvwxb}), for any $i, j \in [F]$, we have $Q_{i, j}(\widehat{F}) = Q_{i, j}(F).$
Thus, we have $\widehat{F} \in \mathscr{F}_{\reg}$,
and for any $i \in [F]$, we have
\begin{equation}
\label{eq:rot-equal2}
\pi_i(\widehat{F}) = \pi_i(F).
\end{equation}
Also, from (\ref{eq:emksr78o7m54}) and (\ref{eq:8ckwdpt4s2b5}), for any $i \in [F]$, we obtain
\begin{eqnarray}
L_i(\widehat{F})
&=& \begin{cases}
 \sum_{s \in \mathcal{S}} |f_i(s)d_{F, \trans_i(s)}| \cdot \mu(s) & \mathrm{if}\,\, \PREF_{F, i} = \{0, 1\}, \\
 \sum_{s \in \mathcal{S}} |\suff(f_i(s)d_{F, \trans_i(s)})| \cdot \mu(s) & \mathrm{if}\,\, \PREF_{F, i} \neq \{0, 1\},\\
\end{cases}\\
&=& \begin{cases}
 L_i(F) + \sum_{s \in \mathcal{S}} d_{F, \trans_i(s)} \cdot \mu(s) & \mathrm{if}\,\, \PREF_{F, i} = \{0, 1\}, \\
 L_i(F) + \sum_{s \in \mathcal{S}} d_{F, \trans_i(s)} \cdot \mu(s) - 1 & \mathrm{if}\,\, \PREF_{F, i} \neq \{0, 1\},\\
\end{cases}\\
&=& \begin{cases}
  L_i(F) +  \sum_{j \in \mathcal{B}} Q_{i, j}(F) & \mathrm{if}\,\, i \not\in \mathcal{B}, \\
  L_i(F) +  \sum_{j \in \mathcal{B}} Q_{i, j}(F) - 1 & \mathrm{if}\,\, i \in \mathcal{B},
\end{cases}
\label{eq:rot-equal3}
\end{eqnarray}
 where $\mathcal{B} := \{i \in [F] : \PREF_{F, i} \neq \{0, 1\}\}$.

Therefore, we have
\begin{eqnarray}
\lefteqn{L(\widehat{F})} \nonumber\\
&=& \sum_{i \in [F]} \pi_i(\widehat{F})L_i(\widehat{F}) \\
&=& \sum_{i \in [F] \setminus \mathcal{B}} \pi_i(\widehat{F})L_i(\widehat{F})
+ \sum_{i \in \mathcal{B}} \pi_i(\widehat{F})L_i(\widehat{F})\\
&=& \sum_{i \in [F] \setminus \mathcal{B}} \pi_i(F)(L_i(F) + \sum_{j \in \mathcal{B}} Q_{i, j}(F)) 
 + \sum_{i \in \mathcal{B}} \pi_i(F)(L_i(F) + \sum_{j \in \mathcal{B}} Q_{i, j}(F) - 1)\\
&=& (\sum_{i \in [F] \setminus \mathcal{B}} \pi_i(F)L_i(F) + \sum_{i \in \mathcal{B}} \pi_i(F)L_i(F)) \nonumber\\
&&+ (\sum_{i \in [F] \setminus \mathcal{B}}\sum_{j \in \mathcal{B}} \pi_i(F)Q_{i, j}(F)
+ \sum_{i \in \mathcal{B}}\sum_{j \in \mathcal{B}} \pi_i(F)Q_{i, j}(F)) - \sum_{i \in \mathcal{B}} \pi_i(F)\\
&=& \sum_{i \in [F]} \pi_i(F)L_i(F) + \sum_{i \in [F]}\sum_{j \in \mathcal{B}} \pi_i(F)Q_{i, j}(F)
- \sum_{j \in \mathcal{B}} \pi_j(F)\\
&=& \sum_{i \in [F]} \pi_i(F)L_i(F) + \sum_{i \in [F]}\sum_{j \in \mathcal{B}} \pi_i(F)Q_{i, j}(F) - \sum_{j \in \mathcal{B}} \sum_{i \in [F]}\pi_i(F)Q_{i, j}(F)\\
&=& \sum_{i \in [F]} \pi_i(F)L_i(F)\\
&=& L(F),
\end{eqnarray}
where the third equality is from (\ref{eq:rot-equal2}) and (\ref{eq:rot-equal3}), and the sixth equality is from (\ref{eq:stationary1}).
\qed

 \begin{example}
Let $\mathcal{S} = \{\mathrm{a}, \mathrm{b},\mathrm{c},\mathrm{d},\mathrm{e}\}$ and $(\mu(\mathrm{a}), \mu(\mathrm{b}), \mu(\mathrm{c}), \mu(\mathrm{d}), \mu(\mathrm{e})) = (0.1, 0.2, 0.2, 0.2, 0.3)$.
Table \ref{tab:rot-equal} shows that $L(F^{(\beta)}), L(F^{(\gamma)})$, and $L(F^{(\delta)})$ are equal.
We can see that the operation of rotation does not change the average codeword length for this example.
\end{example}

\begin{table}[H]
\caption{$L(F^{(\beta)}), L(F^{(\gamma)})$, and $L(F^{(\delta)})$ are equal.}
\label{tab:rot-equal}
\centering

\begin{tabular}{cccccccc}
\hline
$F$ & $L_0(F)$ & $L_1(F)$ & $L_2(F)$ & $\pi_0(F)$ & $\pi_1(F)$ & $\pi_2(F)$ & $L(F)$\\
\hline \hline
$F^{(\beta)}$ & $2.8$ & $3.9$ & $3.7$ & $7/68$ & $26/68$ & $35/68$ & $3.683823\ldots$\\
$F^{(\gamma)}$ & $3.8$ & $3.9$ & $3.5$ & $7/68$ & $26/68$ & $35/68$ & $3.683823\ldots$\\
$F^{(\delta)}$ & $4.4$ & $3.2$ & $3.9$ & $7/68$ & $26/68$ & $35/68$ & $3.683823\ldots$\\
\hline
\end{tabular}
\end{table}
 
Summing up Lemmas \ref{lem:rot-decodable}--\ref{lem:rot-equal}, we obtain the following Lemma \ref{lem:rot-preserve}.
\begin{lemma}
\label{lem:rot-preserve}
For any $F \in \mathscr{F}_{\reg} \cap \mathscr{F}_{\ext} \cap \mathscr{F}_{1\hdec}$, we have $\widehat{F} \in \mathscr{F}_{\reg} \cap \mathscr{F}_{\ext} \cap \mathscr{F}_{1\hdec}$ and $L(\widehat{F}) = L(F)$.
\end{lemma}

\subsection{Any $F \in \mathscr{F}_{\reg} \cap \mathscr{F}_{\ext} \cap \mathscr{F}_{1\hdec}$ Can Be Replaced with Some $F' \in \mathscr{F}_{\reg} \cap\mathscr{F}_{\ext} \cap \mathscr{F}_{0\hdec}$}
\label{subsec:optimality-3}

The goal of this subsection is to prove Lemma \ref{lem:1dec-0dec} that for any $F \in \mathscr{F}_{\reg} \cap \mathscr{F}_{\ext} \cap \mathscr{F}_{1\hdec}$, there exists $F' \in \mathscr{F}_{\reg} \cap\mathscr{F}_{\ext} \cap \mathscr{F}_{0\hdec}$ such that $L(F') = L(F)$.
To prove it, we show Lemma \ref{lem:goodsetTfork} that for any integer $k \geq 0$ and $F \in \mathscr{F}_{\reg} \cap \mathscr{F}_{\ext} \cap \mathscr{F}_{k\hdec}$, there exists $F' \in \mathscr{F}_{\reg} \cap\mathscr{F}_{\ext} \cap \mathscr{F}_{k\hdec} \cap \mathscr{F}_{\fork}$ such that $L(F') = L(F)$,
where $\mathscr{F}_{\fork}$ is defined in the following Definition \ref{def:F_fork}.
Then we prove the desired Lemma \ref{lem:1dec-0dec} arguing $\mathscr{F}_{\reg} \cap \mathscr{F}_{\ext} \cap \mathscr{F}_{1\hdec} \cap \mathscr{F}_{\fork} \subseteq \mathscr{F}_{\reg} \cap \mathscr{F}_{\ext} \cap \mathscr{F}_{0\hdec}$.

\begin{definition}
\label{def:F_fork}
We define $\mathscr{F}_{\fork}$ as $\mathscr{F}_{\fork} := \{F \in \mathscr{F} : {}^\forall i \in [F], \PREF_{F, i} = \{0, 1\}\}$, that is, $\mathscr{F}_{\fork}$ is the set of all code-tuples $F$ such that $\PREF_{F, 0} = \PREF_{F, 1} = \cdots = \PREF_{F, |F|-1} = \{0, 1\}$.
\end{definition}

\begin{example}
Consider the code-tuples of Table \ref{tab:code-tuple}. From Example \ref{ex:pref}, we have $F^{(\alpha)}, F^{(\beta)}, F^{(\gamma)} \not\in \mathscr{F}_{\fork}$ and $F^{(\delta)} \in \mathscr{F}_{\fork}$.
\end{example}

The following Lemma \ref{lem:goodsetTfork} guarantees that we can assume that a given code-tuple is in $\mathscr{F}_{\fork}$ without loss of generality.
\begin{lemma}
\label{lem:goodsetTfork} 
For any integer $k \geq 0$ and $F(f, \trans) \in \mathscr{F}_{\reg} \cap \mathscr{F}_{\ext} \cap \mathscr{F}_{k\hdec}$, there exists $F'(f', \trans') \in \mathscr{F}_{\reg} \cap \mathscr{F}_{\ext} \cap \mathscr{F}_{k\hdec} \cap \mathscr{F}_{\fork}$ such that $L(F') = L(F)$.
\end{lemma}
To prove Lemma \ref{lem:goodsetTfork}, for an integer $k \geq 0$, $F \in \mathscr{F}_{k\hdec} \cap \mathscr{F}_{\ext}$, and $i \in [F]$, we define $l_{F, i}$ as follows.
\begin{equation}
\label{eq:l_F}
l_{F, i} := \min\{|f^{\ast}_i(\pmb{x}) \land f^{\ast}_i(\pmb{x'})| : \pmb{x}, \pmb{x'} \in \mathcal{S}^{\ast}, f^{\ast}_i(\pmb{x}) \not\succpreceq f^{\ast}_i(\pmb{x'})\},
\end{equation}
where $f^{\ast}_i(\pmb{x}) \not\succpreceq f^{\ast}_i(\pmb{x'})$ means $f^{\ast}_i(\pmb{x}) \not\preceq f^{\ast}_i(\pmb{x'})$ and $f^{\ast}_i(\pmb{x}) \not\succeq f^{\ast}_i(\pmb{x'})$,
and $\pmb{x} \land \pmb{x'}$ is the longest common prefix of $\pmb{x}$ and $\pmb{x'}$, that is,
the longest sequence $\pmb{z}$ such that $\pmb{z} \preceq \pmb{x}$ and $\pmb{z} \preceq \pmb{x'}$.

\begin{table}
\caption{$l_{F, i}, i \in [F]$ for the code-tuples $F^{(\beta)}, F^{(\gamma)}$ and $F^{(\delta)}$ in Table \ref{tab:code-tuple}.}
\label{tab:pref}
\centering
\begin{tabular}{cccc}
\hline
$F$ & $l_{F, 0}$ & $l_{F, 1}$ & $l_{F, 2}$\\
\hline \hline
$F^{(\beta)}$ & 1 & 2 & 1\\
$F^{(\gamma)}$ & 0 & 1 & 0\\
$F^{(\delta)}$ & 0 & 0 & 0\\
\hline
\end{tabular}
\end{table}

\begin{example}
Table \ref{tab:pref} shows $l_{F, i}, i \in [F]$ for the code-tuples $F^{(\beta)}, F^{(\gamma)}$, and $F^{(\delta)}$ in Table \ref{tab:code-tuple}.
\end{example}

Note that from (\ref{eq:l_F}) and (\ref{eq:emksr78o7m54}), we obtain
\begin{equation}
\label{eq:mrxwly8sbpbm}
l_{F, i} = 0 \Leftrightarrow \PREF_{F, i} = \{0, 1\} \Leftrightarrow d_{F, i} = 0.
\end{equation}

The following Lemma \ref{lem:fork-exist} guarantees that the right hand side of (\ref{eq:l_F}) is well-defined.
\begin{lemma}
\label{lem:fork-exist}
For any integer $k \geq 0$, $F(f, \trans) \in \mathscr{F}_{k\hdec} \cap \mathscr{F}_{\ext}$, and $i \in [F]$, there exists $\pmb{x}, \pmb{x'} \in \mathcal{S}^{\ast}$ such that $f^{\ast}_i(\pmb{x}) \not\succpreceq f^{\ast}_i(\pmb{x'})$.
\end{lemma}
\emph{Proof of Lemma \ref{lem:fork-exist}}:
We prove by contradiction assuming that there exist an integer $k \geq 0$, $F(f, \trans) \in \mathscr{F}_{k\hdec} \cap \mathscr{F}_{\ext}$, and $i \in [F]$ such that for any $\pmb{x}, \pmb{x'} \in \mathcal{S}^{\ast}$, we have $f^{\ast}_i(\pmb{x}) \preceq f^{\ast}_i(\pmb{x'})$ or $f^{\ast}_i(\pmb{x}) \succeq f^{\ast}_i(\pmb{x'})$.

Choose two distinct symbols $s, s' \in \mathcal{S}$ arbitrarily and assume $f^{\ast}_i(s) \preceq f^{\ast}_i(s')$ without loss of generality.
From $F \in \mathscr{F}_{\ext}$, we can choose $\pmb{x}, \pmb{x'} \in \mathcal{S}^{\ast}$ such that
\begin{equation}
\label{eq:esuq24i9fsl3}
|f^{\ast}_i(s\pmb{x})| = |f^{\ast}_i(s'\pmb{x'})| \geq |f^{\ast}_i(s)| + k.
\end{equation}
Then, from the assumption, we have
\begin{equation}
\label{eq:de87j40akitm}
f^{\ast}_i(s  \pmb{x}) \preceq f^{\ast}_i(s'  \pmb{x'}) \,\,\text{or}\,\, f^{\ast}_i(s  \pmb{x}) \succeq f^{\ast}_i(s'  \pmb{x'}).
\end{equation}
From (\ref{eq:esuq24i9fsl3}) and (\ref{eq:de87j40akitm}),
\begin{equation}
\label{eq:nmkhburgehar}
f^{\ast}_i(s\pmb{x}) = f^{\ast}_i(s'\pmb{x'}) \succeq f^{\ast}_i(s)  \pmb{c},
\end{equation}
for some $\pmb{c} \in \mathcal{C}^k$.
From (\ref{eq:nmkhburgehar}) and $s \preceq s\pmb{x}$, the pair $(s, \pmb{c}) \in \mathcal{S}^1 \times \mathcal{C}^k$ is not $f^{\ast}_i$-negative.
Also, from (\ref{eq:nmkhburgehar}) and $s \not\preceq s'\pmb{x}$, the pair $(s, \pmb{c})$ is not $f^{\ast}_i$-positive.
Consequently, $(s, \pmb{c}) \in \mathcal{S}^1 \times \mathcal{C}^k \subset \mathcal{S}^{\ast} \times \mathcal{C}^k$ is neither $f^{\ast}_i$-positive nor $f^{\ast}_i$-negative.
This conflicts with $F \in \mathscr{F}_{k\hdec}$. 
\qed

Now we state the proof of Lemma \ref{lem:goodsetTfork} as follows.

\noindent
\emph{Proof of Lemma \ref{lem:goodsetTfork}}:
For $t = 0, 1, 2, \ldots$, define $F^{(t)}(f^{(t)}, \trans^{(t)}) \in \mathscr{F}$ as follows.
\begin{equation}
F^{(t)} := 
\begin{cases}
F & \mathrm{if}\,\, t = 0, \\
\widehat{F^{(t-1)}} & \mathrm{if}\,\, t \geq 1,\\
\end{cases}
\end{equation} 
that is, $F^{(t)}$ is the code-tuple obtained by applying $t$ times rotation to $F$.

From $F^{(0)} = F \in \mathscr{F}_{\reg} \cap \mathscr{F}_{\ext} \cap \mathscr{F}_{k\hdec}$ and Lemma \ref{lem:rot-preserve}, for any $t \geq 0$, we have $F^{(t)} \in \mathscr{F}_{\reg} \cap \mathscr{F}_{\ext} \cap \mathscr{F}_{k\hdec}$ and $L(F^{(t)}) = L(F)$.
Therefore, to prove Lemma \ref{lem:goodsetTfork}, it suffices to prove that there exists an integer $\bar{t} \geq 0$ such that $F^{(\bar{t})} \in \mathscr{F}_{\fork}$.
Furthermore, from (\ref{eq:mrxwly8sbpbm}), it suffices to prove that for some integer $\bar{t} \geq 0$, we have
$l_{F^{(\bar{t})}, 0} = l_{F^{(\bar{t})}, 1} = \cdots = l_{F^{(\bar{t})}, |F|-1} = 0$.

Fix $i \in [F]$ and choose $\pmb{x}, \pmb{x'} \in \mathcal{S}^{\ast}$ such that ${f^{\ast}_i}^{(t)}(\pmb{x}) \not\succpreceq {f^{\ast}_i}^{(t)}(\pmb{x'})$ and $|{f^{\ast}_i}^{(t)}(\pmb{x}) \land {f^{\ast}_i}^{(t)}(\pmb{x'})| = l_{F^{(t)}, i}$.

Then we have  ${f^{\ast}_i}^{(t)}(\pmb{x})  d_{F, \trans^{\ast}_i(\pmb{x})} \not\succpreceq {f^{\ast}_i}^{(t)}(\pmb{x'})  d_{F, \trans^{\ast}_i(\pmb{x})}$.
From Lemma \ref{lem:rotation}, we have $d_{F, i}  {f^{\ast}_i}^{(t+1)}(\pmb{x}) \not\succpreceq d_{F, i}  {f^{\ast}_i}^{(t+1)}(\pmb{x'})$.
Hence, we obtain ${f^{\ast}_i}^{(t+1)}(\pmb{x}) \not\succpreceq {f^{\ast}_i}^{(t+1)}(\pmb{x'})$.
Therefore, it holds that
\begin{equation}
l_{F^{(t+1)}, i} \leq |{f^{\ast}_i}^{(t+1)}(\pmb{x}) \land {f^{\ast}_i}^{(t+1)}(\pmb{x'})|.
\end{equation}

Also, from ${f^{\ast}_i}^{(t)}(\pmb{x}) \not\succpreceq {f^{\ast}_i}^{(t)}(\pmb{x'})$ and $|{f^{\ast}_i}^{(t)}(\pmb{x}) \land {f^{\ast}_i}^{(t)}(\pmb{x'})| = l_{F^{(t)}, i}$, we have
\begin{eqnarray}
l_{F^{(t)}, i} &=& |{f^{\ast}_i}^{(t)}(\pmb{x}) \land {f^{\ast}_i}^{(t)}(\pmb{x'})|\\
&=& |({f^{\ast}_i}^{(t)}(\pmb{x})  d_{F^{(t)}, \trans^{\ast}_i(\pmb{x})}) \land ({f^{\ast}_i}^{(t)}(\pmb{x'})  d_{F^{(t)}, \trans^{\ast}_i(\pmb{x'})})|.
\label{eq:wlaxlpiaqdnd}
\end{eqnarray}
From Lemma \ref{lem:rotation},
\begin{eqnarray}
l_{F^{(t)}, i} &=& |(d_{F^{(t)}, i}  {f^{\ast}_i}^{(t+1)}(\pmb{x})) \land (d_{F^{(t)}, i}  {f^{\ast}_i}^{(t+1)}(\pmb{x'}))|\\
&=& |d_{F^{(t)}, i}| + |{f^{\ast}_i}^{(t+1)}(\pmb{x}) \land {f^{\ast}_i}^{(t+1)}(\pmb{x'})|\\
&\geq& |d_{F^{(t)}, i}| + l_{F^{(t+1)}, i}, \label{eq:kj3khfcibbar}
\end{eqnarray}
where the inequality is from (\ref{eq:wlaxlpiaqdnd}).
From (\ref{eq:mrxwly8sbpbm}) and (\ref{eq:kj3khfcibbar}), we have $l_{F^{(t+1)}, i} = 0$ if $l_{F^{(t)}, i} = 0$ and $l_{F^{(t+1)}, i} < l_{F^{(t)}, i}$ if $l_{F^{(t)}, i} > 0$.
Therefore, $l_{F^{(t)}, i} = 0$ for $t \geq l_{F^{(0)}, i}$.
Consequently, $F^{(\bar{t})} \in \mathscr{F}_{\fork}$, where $\bar{t} := \max\{l_{F, 0}, l_{F, 1}, \ldots, \allowbreak l_{F, |F|-1}\}$.
\qed

Now we prove the following Lemma \ref{lem:1dec-0dec} that any $F \in \mathscr{F}_{\reg} \cap \mathscr{F}_{\ext} \cap \mathscr{F}_{1\hdec}$ can be replaced with some $F' \in \mathscr{F}_{\reg} \cap\mathscr{F}_{\ext} \cap \mathscr{F}_{0\hdec}$.
\begin{lemma}
\label{lem:1dec-0dec}
For any $F \in \mathscr{F}_{\reg} \cap \mathscr{F}_{\ext} \cap \mathscr{F}_{1\hdec}$, there exists $F' \in \mathscr{F}_{\reg} \cap\mathscr{F}_{\ext} \cap \mathscr{F}_{0\hdec}$ such that $L(F') = L(F)$.
\end{lemma}

\emph{Proof of Lemma \ref{lem:1dec-0dec}}:
From Lemma \ref{lem:goodsetTfork}, we can choose $F'(f', \trans') \in \mathscr{F}_{\reg} \cap \mathscr{F}_{\ext} \cap  \mathscr{F}_{1\hdec} \cap \mathscr{F}_{\fork}$ such that $L(F') = L(F)$. Now we prove $F' \in \mathscr{F}_{0\hdec}$
by showing that for $i \in [F]$ and $\pmb{x} \in \mathcal{S}^{\ast}$, the pair $(\pmb{x}, \lambda)$ is $f'^{\ast}_i$-positive, that is, for any $i \in [F]$ and $\pmb{x}, \pmb{x'} \in \mathcal{S}^{\ast}$ such that $f'^{\ast}_i(\pmb{x}) \preceq f'^{\ast}_i(\pmb{x'})$, we have $\pmb{x} \preceq \pmb{x'}$.

Choose  $\pmb{x}, \pmb{x'} \in \mathcal{S}^{\ast}$ such that $f'^{\ast}_i(\pmb{x}) \preceq f'^{\ast}_i(\pmb{x'})$ arbitrarily.
Since $F' \in \mathscr{F}_{\fork}$, we have $\PREF_{F', \trans'^{\ast}_i(\pmb{x})} = \{0, 1\}$,
that is, there exists $\pmb{y}_0, \pmb{y}_1 \in \mathcal{S}^{\ast}$ such that $f'^{\ast}_{\trans'^{\ast}_i(\pmb{x})}(\pmb{y}_0) \succeq 0$ and $f'^{\ast}_{\trans'^{\ast}_i(\pmb{x})}(\pmb{y}_1) \succeq 1$.
Hence, we have
$f'^{\ast}_i(\pmb{x}\pmb{y}_0) = f'^{\ast}_i(\pmb{x})  f'^{\ast}_{\trans'^{\ast}_i(\pmb{x})}(\pmb{y}_0) \succeq f'^{\ast}_i(\pmb{x})  0$ and $f'^{\ast}_i(\pmb{x}\pmb{y}_1) = f'^{\ast}_i(\pmb{x})  f'^{\ast}_{\trans'^{\ast}_i(\pmb{x})}(\pmb{y}_1) \succeq f'^{\ast}_i(\pmb{x})  1$, where the equalities are from Lemma \ref{lem:f_T} (i).
Since $\pmb{x} \preceq \pmb{x}  \pmb{y}_0, \pmb{x} \preceq \pmb{x}  \pmb{y}_1$, and $F' \in \mathscr{F}_{1\hdec}$, the pairs $(\pmb{x}, 0), (\pmb{x}, 1)$ are $f'^{\ast}_i$-positive.

From $f'^{\ast}_i(\pmb{x}) \preceq f'^{\ast}_i(\pmb{x'})$ and $F' \in \mathscr{F}_{\ext}$, there exists $c, c' \in \mathcal{C}^1$ and $\pmb{x''} \in \mathcal{S}^{\ast}$ such that $f'^{\ast}_i(\pmb{x})  c \preceq f'^{\ast}_i(\pmb{x'})  c' \preceq f'^{\ast}_i(\pmb{x'}\pmb{x''})$.
Since $(\pmb{x}, 0)$ and $(\pmb{x}, 1)$ are $f'^{\ast}_i$-positive, 
we have $\pmb{x} \preceq \pmb{x'}  \pmb{x''}$.
Therefore, we have either (a) or (b) of the following conditions:
(a) $\pmb{x} \preceq \pmb{x'}$, (b) $\pmb{x} \neq \pmb{x'}$ and $\pmb{x} \succeq \pmb{x'}$.
To complete the proof, it suffices to prove that (a) is true.
Now we prove it by contradiction assuming that (b) is true, that is, there exists $\pmb{z} = z_1z_2\ldots z_{|\pmb{z}|} \in \mathcal{S}^{\ast} \setminus \{\lambda\}$ such that $\pmb{x} = \pmb{x'}  \pmb{z}$.

From $\pmb{x} \succeq \pmb{x'}$, Lemma \ref{lem:f_T} (iii), and $f'^{\ast}_i(\pmb{x}) \preceq f'^{\ast}_i(\pmb{x'})$, we have
\begin{equation}
\label{eq:2umru5qu6w45}
f'^{\ast}_i(\pmb{x'}) = f'^{\ast}_i(\pmb{x}).
\end{equation}

Choose $s \in \mathcal{S} \setminus \{z_{|\pmb{z}|}\}$ and define $\pmb{z'} = z_1z_2\ldots z_{|\pmb{z}|-1}s$.
From $F' \in \mathscr{F}_{\ext}$, we can choose $\pmb{y'} \in \mathcal{S}^{\ast}$ and $c' \in \mathcal{C}$ such that
\begin{equation}
\label{eq:znfnnh1ojiuj}
f'^{\ast}_{\trans'^{\ast}_i(\pmb{x'})}(\pmb{z'}  \pmb{y'}) \succeq c'.
\end{equation}
From (\ref{eq:2umru5qu6w45}) and (\ref{eq:znfnnh1ojiuj}),
\begin{equation}
f'^{\ast}_i(\pmb{x})  c' = f'^{\ast}_i(\pmb{x'})  c' \preceq f'^{\ast}_i(\pmb{x'})  f'^{\ast}_{\trans'^{\ast}_i(\pmb{x'})}(\pmb{z'}  \pmb{y'}) = f'^{\ast}_i(\pmb{x'}  \pmb{z'}  \pmb{y'}),
\end{equation}
where the last equality is from Lemma \ref{lem:f_T} (i).
Since $(\pmb{x}, 0)$ and $(\pmb{x}, 1)$ are $f'^{\ast}_i$-positive (in particular, $(\pmb{x}, c')$ is $f'^{\ast}_i$-positive),
 we have $\pmb{x} \preceq \pmb{x'}  \pmb{z'}  \pmb{y'}$.
From $\pmb{x} = \pmb{x'}  \pmb{z}$, we have 
$\pmb{x'}  \pmb{z} \preceq \pmb{x'}  \pmb{z'}  \pmb{y'}$.
Hence, we obtain $\pmb{z} \preceq \pmb{z'}  \pmb{y'}$.
This conflicts with the definition of $\pmb{z'}$.
\qed

\begin{example}
Consider $F^{(\beta)} \in \mathscr{F}_{\reg} \cap \mathscr{F}_{\ext} \cap \mathscr{F}_{1\hdec}$ of Table \ref{tab:code-tuple}.
Lemma \ref{lem:1dec-0dec} guarantees that there exists $F' \in \mathscr{F}_{\reg} \cap\mathscr{F}_{\ext} \cap \mathscr{F}_{0\hdec}$ such that $L(F') = L(F^{(\beta)})$. Indeed, $F^{(\delta)}$ of Table \ref{tab:code-tuple} satisfies $F^{(\delta)} \in \mathscr{F}_{\reg} \cap\mathscr{F}_{\ext} \cap \mathscr{F}_{0\hdec}$ and $L(F^{(\delta)}) = L(F^{(\beta)})$.
\end{example}

\subsection{Proof of Theorem \ref{thm:1dec-Huff}}
\label{subsec:optimality-4}

Finally, we prove the following Theorem \ref{thm:1dec-Huff} as the main result of this paper.

\setcounter{theorem}{0}
\begin{theorem}
\label{thm:1dec-Huff}
For any $F(f, \trans) \in \mathscr{F}_{\reg} \cap \mathscr{F}_{\ext} \cap \mathscr{F}_{1\hdec}$, we have $L(F) \geq L_{\Huff}$, where $L_{\Huff}$ is the average codeword length of the Huffman code.
\end{theorem}
\emph{Proof of Theorem \ref{thm:1dec-Huff}}:
From Lemma \ref{lem:1dec-0dec}, there exists $F' \in \mathscr{F}_{\reg} \cap\mathscr{F}_{\ext} \cap \mathscr{F}_{0\hdec}$ such that $L(F') = L(F)$. 
Thus, we can assume $F \in \mathscr{F}_{\reg} \cap\mathscr{F}_{\ext} \cap \mathscr{F}_{0\hdec}$ without loss of generality.

Let $a \in \argmin_{i \in [F]} L_i(F)$, and define $F'(f'_0, \trans'_0) \in \mathscr{F}^{(1)}$ as $f'_0 := f_a, \trans'_0 := 0$.
From $F \in \mathscr{F}^{(1)}$, the simultaneous equations (\ref{eq:stationary1}) and (\ref{eq:stationary2}) have the unique solution $\pmb{\pi}(F) = (\pi_0(F)) = (1)$.
Therefore, we have $F \in \mathscr{F}_{\reg}$ and
\begin{equation}
\label{eq:1dec-Huff3}
L(F') = \pi_0(F')L_0(F') = L_a(F) = \sum_{i \in [F]} \pi_i(F) L_a(F) \leq \sum_{i \in [F]} \pi_i(F) L_i(F) = L(F),
\end{equation}
where the inequality is from $a \in \argmin_{i \in [F]} L_i(F)$.

From $F \in \mathscr{F}_{0\hdec}$ and Lemma \ref{lem:0dec-prefixfree}, 
for any $i \in [F]$, $f_i$ is prefix-free. 
In particular, $f'_0 = f_a$ is prefix-free.
Therefore, from Lemma \ref{lem:0dec-prefixfree},
we have $F' \in \mathscr{F}_{0\hdec}.$
Hence, from Lemma \ref{lem:0dec-unique} (ii), the code table $f'^{\ast}_0$ is injective, that is, $f'^{\ast}_0$ is uniquely decodable code with a single code table.
Therefore, from McMillan's Theorem\cite{McMillan1956}, we have
\begin{equation}
\label{eq:1dec-Huff4}
L(F') \geq L_{\Huff}.
\end{equation}
From (\ref{eq:1dec-Huff3}) and (\ref{eq:1dec-Huff4}), we obtain $L(F) \geq L_{\Huff}$.
\qed

\section{Conclusion}
\label{sec:conclusion}

This paper considered a data compression system for an i.i.d.\ source. 
We discussed the optimality of Huffman code in the class of $1$-bit delay decodable code with a finite number of code tables.
First, we introduced a code-tuple as a model of a time-variant encoder with a finite number of code tables.
Next, we define the $k$-bit delay decodable code-tuples class for $k = 0, 1, 2, \ldots$.
Then we proved Theorem \ref{thm:1dec-Huff}, which claims that Huffman code achieves the optimal average codeword length in the class of $1$-bit delay decodable code-tuples.

\section*{Appendix A}

\subsection*{Proof of Lemma \ref{lem:rot-prefix}}

To state the proof of Lemma \ref{lem:rot-prefix}, first we prove the following Lemma \ref{lem:pref-prefix}.
\begin{lemma}
\label{lem:pref-prefix}
For $F(f, \trans) \in \mathscr{F}, i \in [F]$, and $\pmb{x}, \pmb{x'} \in \mathcal{S}^{\ast}$,
if $\pmb{x'} \preceq \pmb{x}$ and $f^{\ast}_i(\pmb{x}) = f^{\ast}_i(\pmb{x'})$, then $\PREF_{F, \trans^{\ast}_i(\pmb{x'})} \supseteq \PREF_{F, \trans^{\ast}_i(\pmb{x})}$.
\end{lemma}
\emph{Proof of Lemma \ref{lem:pref-prefix}}:
Let $c \in \PREF_{F, \trans^{\ast}_i(\pmb{x})}$. Then there exists $\pmb{y} \in \mathcal{S}^{\ast}$ such that $f^{\ast}_{\trans^{\ast}_i(\pmb{x})}(\pmb{y}) \succeq c$, and we have $f^{\ast}_i(\pmb{x})  c \preceq f^{\ast}_i(\pmb{x})  f^{\ast}_{\trans^{\ast}_i(\pmb{x})}(\pmb{y})$.
From the assumption that $f^{\ast}_i(\pmb{x}) = f^{\ast}_i(\pmb{x'})$, we have $f^{\ast}_i(\pmb{x'})  c \preceq f^{\ast}_i(\pmb{x})  f^{\ast}_{\trans_i^{\ast}(\pmb{x})}(\pmb{y})$.
Let $\pmb{x} = \pmb{x'}  \pmb{z}$. Then we have 
$f^{\ast}_i(\pmb{x'})  c \preceq f^{\ast}_i(\pmb{x'}  \pmb{z}) + f^{\ast}_{\trans^{\ast}_i(\pmb{x'}\pmb{z})}(\pmb{y})$.
From Lemma \ref{lem:f_T} (i), we have
$f^{\ast}_i(\pmb{x'})  c \preceq f^{\ast}_i(\pmb{x'})  f^{\ast}_{\trans^{\ast}_i(\pmb{x'})}(\pmb{z})  f^{\ast}_{\trans^{\ast}_i(\pmb{x'}\pmb{z})}(\pmb{y})$.
Thus, it holds that $c \preceq f^{\ast}_{\trans_i^{\ast}(\pmb{x'})}(\pmb{z})  f^{\ast}_{\trans_i^{\ast}(\pmb{x'}\pmb{z})}(\pmb{y}) = f^{\ast}_{\trans_i^{\ast}(\pmb{x'})}(\pmb{z}  \pmb{y})$ from Lemma \ref{lem:f_T} (i) and Lemma \ref{lem:f_T} (ii),
Hence, we obtain $c \in \PREF_{F, \trans^{\ast}_i(\pmb{x'})}$.
\qed

\noindent
\emph{Proof of Lemma \ref{lem:rot-prefix}}:
Let $(k, i, \pmb{x}, \pmb{x'})$ be a tuple satisfying all of the conditions (i)--(iii), and we lead a contradiction.

From $\pmb{x'} \preceq \pmb{x}$ of the condition (iii) and Lemma \ref{lem:f_T} (iii), we have
\begin{equation}
\label{eq:rot-prefix}
f^{\ast}_i(\pmb{x'}) \preceq f^{\ast}_i(\pmb{x}).
\end{equation}

From the condition (ii) $|\widehat{f}^{\ast}_i(\pmb{x})| + k \leq |\widehat{f}^{\ast}_i(\pmb{x'})|$,
we have
$|d_{F, i}| + |\widehat{f}^{\ast}_i(\pmb{x})| + k \leq |d_{F, i}| + |\widehat{f}^{\ast}_i(\pmb{x'})|$.
Thus,
$|d_{F, i}  \widehat{f}^{\ast}_i(\pmb{x})| + k \leq |d_{F, i}  \widehat{f}^{\ast}_i(\pmb{x'})|$.
From Lemma \ref{lem:rotation}, we have
$|f^{\ast}_i(\pmb{x})  d_{F, \trans^{\ast}_i(\pmb{x})}| + k \leq |f^{\ast}_i(\pmb{x'})  d_{F, \trans^{\ast}_i(\pmb{x'})}|$.
Consequently,
\begin{equation}
\label{eq:rot-prefix2}
|f^{\ast}_i(\pmb{x'})| \geq |f^{\ast}_i(\pmb{x})| + |d_{F, \trans^{\ast}_i(\pmb{x})}| + k - |d_{F, \trans^{\ast}_i(\pmb{x'})}|.
\end{equation}
From (\ref{eq:rot-prefix}),
\begin{equation}
|d_{F, \trans^{\ast}_i(\pmb{x})}| + k - |d_{F, \trans^{\ast}_i(\pmb{x'})}| \leq 0.
\end{equation}
Since $0 \leq |d_{F, \trans^{\ast}_i(\pmb{x})}| \leq 1$ and $0 \leq |d_{F, \trans^{\ast}_i(\pmb{x'})}| \leq 1$,
the following two cases are possible:
(i) $|d_{F, \trans^{\ast}(\pmb{x})}| + k - |d_{F, \trans^{\ast}(\pmb{x'})}| = 0$ and
(ii) $|d_{F, \trans^{\ast}(\pmb{x})}| + k - |d_{F, \trans^{\ast}(\pmb{x'})}| = -1$.

\emph{(i) the case $|d_{F, \trans^{\ast}_i(\pmb{x})}| + k - |d_{F, \trans^{\ast}_i(\pmb{x'})}| = 0$}:
From (\ref{eq:rot-prefix2}), 
\begin{equation}
\label{eq:rot-prefix3}
|f^{\ast}_i(\pmb{x'})| \geq |f^{\ast}_i(\pmb{x})|.
\end{equation}
From (\ref{eq:rot-prefix}) and (\ref{eq:rot-prefix3}),
\begin{equation}
\label{eq:rot-prefix4}
f^{\ast}_i(\pmb{x}) = f^{\ast}_i(\pmb{x'}).
\end{equation}

From Lemma \ref{lem:pref-prefix}, $\PREF_{F, \trans^{\ast}_i(\pmb{x'})} \supseteq \PREF_{F, \trans^{\ast}_i(\pmb{x})}$. In particular, 
\begin{equation}
\PREF_{F, \trans^{\ast}_i(\pmb{x})} = \{0, 1\} \Rightarrow \PREF_{F, \trans^{\ast}_i(\pmb{x'})} = \{0, 1\}.
\end{equation}
From (\ref{eq:emksr78o7m54}),
\begin{equation}
\label{eq:rot-prefix6}
|d_{F, \trans^{\ast}_i(\pmb{x})}| = 0 \Rightarrow |d_{F, \trans^{\ast}_i(\pmb{x'})}| = 0.
\end{equation}
From (\ref{eq:rot-prefix6}) and $|d_{F, \trans^{\ast}_i(\pmb{x})}| + k - |d_{F, \trans^{\ast}_i(\pmb{x'})}| = 0$, we have $k = 0$.
Hence, from the condition (i), we obtain $F \in \mathscr{F}_{0\hdec}$.
Therefore, from (\ref{eq:rot-prefix4}) and Lemma \ref{lem:0dec-unique} (ii), we obtain
$\pmb{x} = \pmb{x'}$,
which conflicts with $\pmb{x} \neq \pmb{x'}$ of the condition (ii).

\emph{(ii) the case $|d_{F, \trans^{\ast}_i(\pmb{x})}| + k - |d_{F, \trans^{\ast}_i(\pmb{x'})}| = -1$}:
We have $|d_{F, \trans^{\ast}_i(\pmb{x})}| =  k  = 0$ and $|d_{F, \trans^{\ast}_i(\pmb{x'})}| = 1$.
From the condition (i) and $k = 0$, we have
\begin{equation}
\label{eq:3id2m4ardshs}
F \in \mathscr{F}_{0\hdec}.
\end{equation}

From (\ref{eq:rot-prefix}),
\begin{equation}
\label{eq:h2x5vljfqubd}
|f^{\ast}_i(\pmb{x'})| \leq |f^{\ast}_i(\pmb{x})|.
\end{equation}
From (\ref{eq:rot-prefix2}) and $|d_{F, \trans^{\ast}_i(\pmb{x})}| + k - |d_{F, \trans^{\ast}_i(\pmb{x'})}| = -1$, 
\begin{equation}
\label{eq:u8b3a80a1drg}
|f^{\ast}_i(\pmb{x})| \leq |f^{\ast}_i(\pmb{x'})|+1.
\end{equation}
From (\ref{eq:h2x5vljfqubd}) and (\ref{eq:u8b3a80a1drg}), 
we have either $|f^{\ast}_i(\pmb{x})| = |f^{\ast}_i(\pmb{x'})|$ or $|f^{\ast}_i(\pmb{x})| + 1 = |f^{\ast}_i(\pmb{x'})|$.
If we assume $|f^{\ast}_i(\pmb{x})| = |f^{\ast}_i(\pmb{x'})|$, then $f^{\ast}_i(\pmb{x}) = f^{\ast}_i(\pmb{x'})$ holds from (\ref{eq:rot-prefix}).
Then from (\ref{eq:3id2m4ardshs}) and Lemma \ref{lem:0dec-unique} (ii), we obtain $\pmb{x} = \pmb{x'}$, which conflicts with $\pmb{x} \neq \pmb{x'}$ of the condition (ii).
Hence, we have
\begin{equation}
\label{eq:ix8cdo8x3ssw}
|f^{\ast}_i(\pmb{x})| = |f^{\ast}_i(\pmb{x'})|+1.
\end{equation}
From the condition (ii) $\pmb{x'} \preceq \pmb{x}$ and $\pmb{x} \neq \pmb{x'}$, there exists $\pmb{z} = z_1z_2\ldots z_{|\pmb{z}|} \in \mathcal{S}^{\ast} \setminus \{\lambda\}$ such that $\pmb{x} = \pmb{x'} + \pmb{z}$.
For such $\pmb{z}$, from (\ref{eq:ix8cdo8x3ssw}), we have
$|f^{\ast}_i(\pmb{x'}  \pmb{z})| = |f^{\ast}_i(\pmb{x'})|+1$.
Then from Lemma \ref{lem:f_T} (i), we have $|f^{\ast}_i(\pmb{x'})| + |f^{\ast}_{\trans^{\ast}_i(\pmb{x'})}(\pmb{z})| = |f^{\ast}_i(\pmb{x'})|+1$.
Thus, we obtain
\begin{eqnarray}
|f^{\ast}_{\trans^{\ast}_i(\pmb{x'})}(\pmb{z})| = 1. \label{eq:wm04p3uaf80s}
\end{eqnarray}

Choose $s \in \mathcal{S} \setminus \{z_{|\pmb{z}|}\}$ and define $\pmb{z'} := z_1z_2\ldots z_{|\pmb{z}|-1}s$.
From $F \in \mathscr{F}_{\ext}$, we can choose $\pmb{y} \in \mathcal{S}^{\ast}$ such that
\begin{equation}
\label{eq:gktnq1mbkhep}
|f^{\ast}_{\trans^{\ast}_i(\pmb{x'})}(\pmb{z'}  \pmb{y})| \geq 1.
\end{equation}

From $|d_{F, \trans^{\ast}_i(\pmb{x'})}| = 1$ (i.e., $\PREF_{F, \trans^{\ast}_i(\pmb{x'})} = \{d_{F, \trans^{\ast}_i(\pmb{x'})}\}$), (\ref{eq:wm04p3uaf80s}) and (\ref{eq:gktnq1mbkhep}), we have
\begin{equation}
\label{eq:chgw0sp4r19n}
f^{\ast}_{\trans^{\ast}_i(\pmb{x'})}(\pmb{z}) = d_{F, \trans^{\ast}_i(\pmb{x'})},
\end{equation}
and 
\begin{equation}
\label{eq:fbzmqseyzust}
f^{\ast}_{\trans^{\ast}_i(\pmb{x'})}(\pmb{z'}  \pmb{y}) \succeq d_{F, \trans^{\ast}_i(\pmb{x'})}.
\end{equation}
Therefore,
\begin{equation}
\label{eq:rot-prefix5}
f^{\ast}_i(\pmb{x'}  \pmb{z}) = f^{\ast}_i(\pmb{x'})  f^{\ast}_{\trans^{\ast}_i(\pmb{x'})}(\pmb{z})
= f^{\ast}_i(\pmb{x'})  d_{F, \trans^{\ast}_i(\pmb{x'})}
\preceq f^{\ast}_i(\pmb{x'})  f^{\ast}_{\trans^{\ast}_i(\pmb{x'})}(\pmb{z'}  \pmb{y})
= f^{\ast}_i(\pmb{x'}  \pmb{z'}  \pmb{y}),
\end{equation}
where the first equality is from Lemma \ref{lem:f_T} (i), the second equality is from (\ref{eq:chgw0sp4r19n}),
the relation `$\preceq$' is from (\ref{eq:fbzmqseyzust}), and the last equality is from Lemma \ref{lem:f_T} (i).
From (\ref{eq:3id2m4ardshs}) and Lemma \ref{lem:0dec-unique} (i), we have
$\pmb{x'}  \pmb{z} \preceq \pmb{x'}  \pmb{z'}  \pmb{y}.$
Thus, we obtain $\pmb{z} \preceq \pmb{z'}  \pmb{y}$.
This conflicts with the definition of $\pmb{z'}$.
\qed

\section*{Appendix B}
\subsection*{List of Notations}
\begin{longtable}{cp{14.5cm}}
  $\mathcal{A} \times \mathcal{B}$ & $\{(a, b) : a \in \mathcal{A}, b \in \mathcal{B}\}$, defined at the beginning of Section \ref{sec:preliminary}. \\
  $|\mathcal{A}|$ & the cardinality of a set $\mathcal{A}$, defined at the beginning of Section \ref{sec:preliminary}. \\
  $\mathcal{A}^k$ & the set of all sequences of length $k$ over a set $\mathcal{A}$, defined at the beginning of Section \ref{sec:preliminary}. \\
  $\mathcal{A}^{\ast}$ & the set of all sequences of finite length over a set $\mathcal{A}$, defined at the beginning of Section \ref{sec:preliminary}.\\
  $\mathcal{C}$ & the coding alphabet $\mathcal{C} = \{0, 1\}$, defined in Section \ref{sec:introduction} (above Fig. \ref{fig:system}). \\
  $d_{F, i}$ & defined in (\ref{eq:emksr78o7m54}). \\
  $f^{\ast}_i$ & defined in (\ref{eq:fstar}). \\
  $F$ & simplified notation of a code-tuple $F(f_i, \trans_i : i \in [m])$, also written as $F(f, \trans)$, defined below Definition \ref{def:treepair}.\\
  $|F|$ & the number of code tables of $F$, defined below Definition \ref{def:treepair}. \\
  $[F]$ & simplified notation of $[|F|] = \{0, 1, 2, \ldots, |F|-1\}$, defined below Definition \ref{def:treepair}.\\
  $\widehat{F}$ & the code-tuple obtained by applying rotation to $F$, defined in Definition \ref{def:rotation}. \\
  $\mathscr{F}^{(m)}$ & the set of all $m$-code-tuples, defined after Definition \ref{def:treepair}.\\
  $\mathscr{F}$ & the set of all code-tuples, defined after Definition \ref{def:treepair}.\\
  $\mathscr{F}_{\ext}$ & defined in Definition \ref{def:F_ext}. \\
  $\mathscr{F}_{\fork}$ & defined in Definition \ref{def:F_fork}. \\
  $\mathscr{F}_{k\hdec}$ & the set of all $k$-bit delay decodable code-tuples, defined in Definition \ref{def:k-bitdelay}. \\
  $\mathscr{F}_{\reg}$ & the set of all regular code-tuples, defined in Definition \ref{def:regular}. \\
  $l_{F, i}$ & defined in (\ref{eq:l_F}). \\
  $L(F)$ & the average codeword length of a code-tuple $F$, defined in Definition \ref{def:evaluation}. \\
  $L_i(F)$ & the average codeword length of the $i$-th code table of $F$, defined in Definition \ref{def:evaluation}. \\
  $[m]$ & $\{0, 1, 2, \ldots, m-1\}$, defined at the beginning of Subsection \ref{subsec:treepair}. \\
  $\mathcal{P}_{F, i}$ & the set of all $c \in \mathcal{C}$ which is the first bit of $f^{\ast}_i(\pmb{x})$ for some $\pmb{x} \in \mathcal{S}^{\ast}$, defined in Definition \ref{def:pref}.\\
  $Q(F)$ & the transition probability matrix, defined in Definition \ref{def:transprobability}.\\
  $Q_{i, j}(F)$ & the transition probability, defined in Definition \ref{def:transprobability}.\\
  $\mathcal{S}$ & the source alphabet, defined at the beginning of Section \ref{sec:introduction}. \\
  $\pmb{x} \land \pmb{y}$ & the longest common prefix of $\pmb{x}$ and $\pmb{y}$, defined after Lemma \ref{lem:goodsetTfork}.\\
  $\pmb{x} \preceq \pmb{y}$ & \pmb{x} is a prefix of \pmb{y}, defined at the beginning of Section \ref{sec:preliminary}.\\
  $\pmb{x} \not\succpreceq \pmb{y}$ & \pmb{x} is not a prefix of \pmb{y}, and \pmb{y} is not a prefix of \pmb{x}, defined after Lemma \ref{lem:goodsetTfork}.\\
  $\suff(\pmb{x})$ & the sequence obtained by deleting the first letter of $\pmb{x}$, defined at the beginning of Section \ref{sec:preliminary}. \\
  $|\pmb{x}|$ & the length of a sequence $\pmb{x}$, defined at the beginning of Section \ref{sec:preliminary}.\\
  $\lambda$ & the empty sequence, defined at the beginning of Section \ref{sec:preliminary}.\\
  $\mu(s)$ & the probability of occurrence of symbol $s$, defined at the beginning of Subsection \ref{subsec:evaluation}. \\
  $\pmb{\pi}(F)$ & defined in Definition \ref{def:regular}.\\
  $\sigma$ & the alphabet size, defined at the beginning of Section \ref{sec:introduction}.\\
  $\trans^{\ast}_i$ & defined in (\ref{eq:tstar}). \\
\end{longtable}




%


\end{document}